\documentclass[sn-chicago]{sn-jnl}

\jyear{2021}%

\theoremstyle{thmstyleone}%
\usepackage{comment}

\theoremstyle{thmstyletwo}%
\newtheorem{remark}{Remark}%

\theoremstyle{thmstylethree}%

\renewcommand{\vec}[1]{{\ensuremath{\boldsymbol{\mathrm #1}}}}
\newcommand{\ten}[1]{\ensuremath{\boldsymbol{\mathsf{#1}}}}
\newcommand{\tI}{{\ensuremath{\ten I}}}
\newcommand{\vdot}{\boldsymbol{\mathsf{\ensuremath\cdot}}}

\newcommand{\del}{\ensuremath{\nabla}}
\newcommand{\deld}{\ensuremath{\del\vdot}}
\newcommand{\lrp}[1]{\left( #1 \right)}

\newcommand{\BJ}{{\mathrm{BJ}}}
\newcommand{\bl}{{\mathrm{bl}}}
\newcommand{\ps}{{\mathrm{ps}}}
\newcommand{\mi}{{\mathrm{mi}}}
\newcommand{\ma}{{\mathrm{ma}}}
\newcommand{\aopt}{\alpha_\BJ^\text{opt}}
\newcommand{\FF}{{\mathrm{ff}}}
\newcommand{\PM}{{\mathrm{pm}}}
\newcommand{\RR}{\mathbb{R}}

\newcommand{\Zp}{\ensuremath{Z^+}}
\newcommand{\Zm}{\ensuremath{Z^-}}
\renewcommand{\sec}{Section}

\newcommand{\algo}{Alg.}

\usepackage{stmaryrd}

\usepackage{soul}

\newcommand{\todo}[1]{{\color{red!80!black} {ToDo: #1}}}

\begin{document}

\title[Modified Beavers--Joseph condition for arbitrary flows]{A modification of the Beavers--Joseph condition for arbitrary flows to the fluid--porous interface}


\author[1]{\fnm{Paula} \sur{Strohbeck}}\email{paula.strohbeck@ians.uni-stuttgart.de}

\author[1]{\fnm{Elissa} \sur{Eggenweiler}}\email{elissa.eggenweiler@ians.uni-stuttgart.de}

\author*[1]{\fnm{Iryna} \sur{Rybak}}\email{iryna.rybak@ians.uni-stuttgart.de}

\affil[1]{\orgdiv{Institute of Applied Analysis and Numerical Simulation}, \orgname{University of Stuttgart}, \orgaddress{\street{Pfaffenwaldring 57}, \city{Stuttgart}, \postcode{70569}, \country{Germany}}}

\abstract{
Physically consistent coupling conditions at the fluid--porous interface with correctly determined effective parameters are necessary for accurate modeling and simulation of various applications. To describe single-fluid-phase flows in coupled free-flow and porous-medium systems, the Stokes/Darcy equations are typically used together with the conservation of mass across the interface, the balance of normal forces and the Beavers--Joseph condition on the tangential velocity. The latter condition is suitable for flows parallel to the interface but not applicable for arbitrary flow directions. Moreover, the value of the Beavers--Joseph slip coefficient is uncertain. In the literature, it is routinely set equal to one that is not correct for many applications, even if the flow is parallel to the porous layer. In this paper, we reformulate the generalized interface condition on the tangential velocity component, recently developed for arbitrary flows in Stokes/Darcy systems, such that it has the same analytical form as the Beavers--Joseph condition. We compute the effective coefficients appearing in this modified condition using theory of homogenization with boundary layers. We demonstrate that the modified Beavers--Joseph condition is applicable for arbitrary flow directions to the fluid--porous interface. In addition, we propose an efficient two-level numerical algorithm based on simulated annealing to compute the optimal Beavers--Joseph parameter.

}




\maketitle

\subsubsection*{Article Highlights}
\begin{itemize}
    \item A modification of the Beavers--Joseph condition is proposed based on recently developed generalized coupling conditions.
    \vspace{0.5ex}
    \item The Beavers–Joseph parameter can be found only for unidirectional flows.
    \vspace{0.5ex}
    \item An efficient numerical algorithm to determine the optimal Beavers–Joseph parameter is developed.
\end{itemize}

\section{Introduction}
\label{sec:introduction}

Coupled porous-medium and free-flow systems appear in a variety of environmental and technical applications such as interactions between surface and subsurface water, geothermal problems and industrial filtration processes~\citep{Beaude_etal_19, Cimolin_Discacciati_13, Reuter_etal_19}. From the macroscale perspective, fluid flow in the free-flow domain and through the porous medium is described by two different sets of equations, which need to be coupled at the fluid--porous interface~\citep{Angot_etal_17, Discacciati_Miglio_Quarteroni_02, Jaeger_Mikelic_09, Lacis_Bagheri_17, Beavers_Joseph_67}. Fluid flows through such coupled systems are highly interface driven, therefore, the correct choice of coupling conditions and effective model parameters is crucial for physically consistent modeling and accurate numerical simulation of applications. 

Depending on the flow problem, different mathematical models and different sets of coupling conditions at the fluid--porous interface are applied, e.g.,~\citep{Dawson_08, Angot_etal_17, Jaeger_Mikelic_09, Lacis_Bagheri_17,  Carraro_etal_15, Sochala_Ern_Piperno_09, Eggenweiler_Rybak_20, Discacciati_Quarteroni_09, GiraultRiviere}. Most often, the Navier--Stokes equations or, in case of low Reynolds numbers, the Stokes equations are used to describe fluid flow in the free-flow region while in the porous-medium domain Darcy's law or one of its generalizations is considered. Besides the Stokes equations, there are several other simplifications of the Navier--Stokes equations, which can be used to model surface flows, such as the shallow water equations, kinematic or diffusive wave equations~\citep{Magiera_etal_15, Sochala_Ern_Piperno_09, Dawson_08}. For higher Reynolds numbers, the Forchheimer equation is used in the porous medium and coupled to the free-flow model~\citep{Angot_etal_17, Cimolin_Discacciati_13}. To describe two-fluid-phase flows in the subsurface, the Richards equation or two-phase Darcy's law is applied~\citep{Mosthaf_Baber_etal_11, Beaude_etal_19, Rybak_etal_15}.

Based on the choice of the underlying flow models in the free-flow region and in the porous medium, different sets of coupling conditions at the fluid--porous interface are available in the literature. Interface conditions for the Navier--Stokes/Darcy--Forchheimer model are developed in~\citep{Cimolin_Discacciati_13,Amara_09, AlazmiVafai,Angot_etal_20}. Coupling conditions for the Stokes equations and multiphase Darcy's law together with the transport of chemical species and energy are proposed in~\citep{Mosthaf_Baber_etal_11}. Coupling strategies for the Stokes/Darcy--Brinkman equations are derived in~\citep{Angot_etal_17}. The most widely studied problem in the literature, both from the modeling and numerical point of view, is the Stokes/Darcy problem describing single-fluid-phase flows in coupled free-flow and porous-medium systems. For this model, different coupling strategies exist,  e.g.,~\citep{Beavers_Joseph_67,Angot_etal_17,Lacis_Bagheri_17,Jaeger_Mikelic_00,Eggenweiler_Rybak_MMS20,OchoaTapia_Whitaker_95}. However, the classical set of coupling conditions, which comprises the conservation of mass across the interface, the balance of normal forces and the Beavers--Joseph condition on the tangential velocity~\citep{Beavers_Joseph_67}, is routinely used. Often, the Saffman simplification of the Beavers--Joseph condition neglecting the porous-medium velocity at the interface is considered~\citep{Saffman}. An extension of the Beavers--Joseph coupling condition with the symmetrized viscous stress tensor was postulated by~\cite{Jones_73}.

Originally, the Beavers--Joseph coupling condition was proposed for flows parallel to the fluid--porous interface. Nevertheless, this condition and also its variants obtained by \cite{Saffman} and~\cite{Jones_73} are often used for non-parallel flows to the porous layer, e.g.,~\citep{HanspalWaghodeNassehiWakeman, Discacciati_Gerardo-Giorda_18}. In~\citep{Eggenweiler_Rybak_20}, it is shown that the Beavers--Joseph condition is not applicable for arbitrary flow directions to the interface that is the case for e.g. filtration problems. Besides the unsuitability of the Beavers--Joseph coupling condition~\eqref{eq:IC-BJ} for non-parallel flows to the interface, the slip coefficient $\alpha_\BJ$ is uncertain and needs to be determined for every flow problem. Although the question concerning the correct value of the Beavers--Joseph parameter is known for decades, to the best of the authors knowledge, no systematic study on determination of this parameter has been published. In case of cylindrical grains and parallel flows to the porous bed, the values of the Beavers--Joseph slip coefficient were obtained in~\citep{Mierzwiczak_19}. In the literature, $\alpha_\BJ=1$ is typically used even if this is not the correct choice~\citep{Yang_etal_19,Eggenweiler_Rybak_20,Rybak_etal_19,Lacis_etal_20}. However, the correct value of the Beavers--Joseph parameter is essential for accurate numerical simulations of applications, where the Beavers--Joseph condition is applicable, e.g., for microfluidic experiments~\citep{Weishaupt_2019}. 

Alternative coupling concepts for the Stokes/Darcy problem exist in the literature. However, some of them contain unknown model parameters, which still need to be determined before the interface conditions can be used in numerical simulations of applications, e.g.~\citep{Angot_etal_17,Angot_etal_20, OchoaTapia_Whitaker_95}. Other alternative coupling concepts, where the effective coefficients can be computed based on the pore geometry, are either derived only for unidirectional flows parallel or perpendicular to the porous layer~\citep{Jaeger_Mikelic_00, Jaeger_Mikelic_09,Carraro_etal_15} or they are not validated for arbitrary flow directions~\citep{Lacis_Bagheri_17,Lacis_etal_20,Zampogna_Bottaro_16}. An exception is the set of generalized interface conditions proposed in~\citep{Eggenweiler_Rybak_MMS20}, which is valid for arbitrary flows in coupled free-flow and porous-medium systems. This set of coupling conditions comprises the conservation of mass across the interface~\eqref{eq:IC-mass} or~\eqref{eq:ER-mass}, the classical balance of normal forces~\eqref{eq:IC-momentum} for isotropic porous media and its extension \eqref{eq:ER-momentum} in case of anisotropic media, and a generalization \eqref{eq:ER-tangential} of the Beavers--Joseph condition~\eqref{eq:IC-BJ} on the tangential velocity. Condition~\eqref{eq:ER-tangential} contains an interfacial porous-medium velocity instead of the Darcy velocity, which appears in the Beavers--Joseph condition~\eqref{eq:IC-BJ}, and the boundary layer coefficient $\vec N^\bl$ instead of the Beavers--Joseph slip parameter $\alpha_\BJ$. The goal of this paper is to reformulate the generalized interface condition~\eqref{eq:ER-tangential} in the same analytical form as the traditionally used Beavers--Joseph condition~\eqref{eq:IC-BJ} and to provide the means to compute effective coefficients appearing in the modified condition. In this way, other researchers can use their already developed software for coupled flow problems based on the classical interface conditions and only adjust the effective model parameters.

In this paper, we study two different flow scenarios: a pressure-driven flow parallel to the porous layer and a general filtration problem with arbitrary flow directions to the interface. We formulate the generalized interface condition~\eqref{eq:ER-tangential} on the tangential velocity component in the form of the Beavers--Joseph condition~\eqref{eq:IC-BJ} and determine the effective model parameters for various porous-medium geometrical configurations (isotropic, orthotropic, anisotropic). To compute the effective coefficients for periodic media, the theory of homogenization and boundary layers~\citep{Jaeger_Mikelic_96,Hornung_97} is applied. As an alternative approach, we propose an efficient two-level numerical algorithm, which is valid also for non-periodic porous structures. We demonstrate that the Beavers--Joseph slip coefficient $\alpha_\BJ$ can be found only for unidirectional flows to the interface and compute the optimal value of this parameter for all considered pore geometries. We show that $\alpha_\BJ$ cannot be fitted for arbitrary flow directions, and in this case modification~\eqref{eq:IC-BJ-mod} has to be used. We perform model validation by comparison of pore-scale resolved to macroscale numerical simulation results.

The paper is organized as follows. In~\sec~\ref{sec:models}, we describe the pore-scale resolved and the macroscale flow models including two sets of coupling conditions on the fluid--porous interface: the classical and the modified ones. In~\sec~\ref{sec:parameters}, we provide the approach to compute  permeability and boundary layer constants appearing in the interface conditions and propose an efficient two-level numerical algorithm to determine the optimal value of the Beavers--Joseph slip coefficient. The detailed numerical study of  different coupled flow problems is conducted in~\sec~\ref{sec:results} and the conclusions follow in~\sec~\ref{sec:conclusions}.

\section{Mathematical models}
\label{sec:models}

In this section, we describe the flow system of interest and present the microscale (pore-scale resolved) and the macroscale flow models. The microscale model is used for the determination of the optimal Beavers--Joseph slip coefficient for unidirectional flows, for the computation of the effective coefficients in the modified conditions, and for the validation of interface conditions. The macroscale model consists of the Stokes/Darcy equations with the classical set of coupling conditions and the  modified ones, respectively. 

\subsection{Assumptions on geometry and flow}
\label{sec:assumptions}
\begin{figure}[!ht]
\centerline{\includegraphics[height=0.35\textwidth]{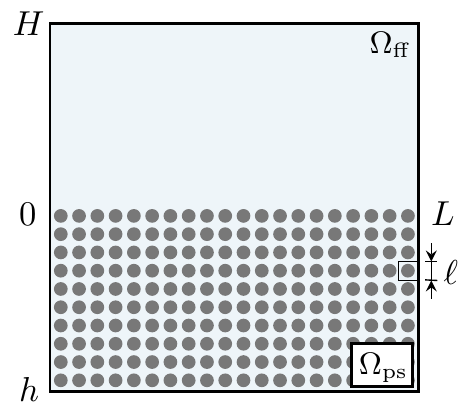}\qquad \quad
\includegraphics[height=0.35\textwidth]{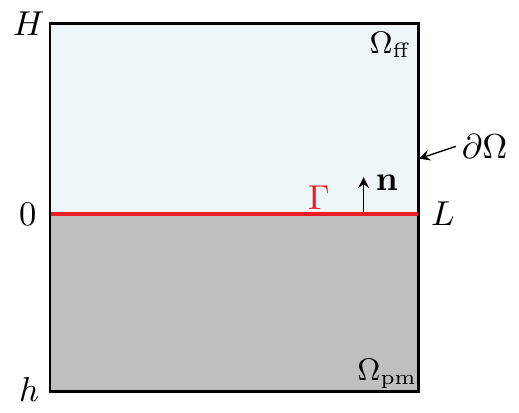}}
\caption{Flow system description at the microscale (left) and macroscale (right) 
}
\label{fig:domain}
\end{figure}

We consider steady-state fluid flow in the coupled domain $\Omega = \Omega_\FF \cup \Omega_\PM$ consisting of the free-flow region $\Omega_\FF \subset \RR^2$ and the porous medium $\Omega_{\text{pm}} \subset \RR^2$. The porous layer contains periodically distributed solid inclusions that allows to compute the permeability and the effective boundary layer coefficients in the modified interface conditions using the theory of homogenization. We assume the separation of scales, i.e., $\ell \ll L$, where $\ell$ denotes the characteristic pore size and $L$ is the length of the domain (Fig.~\ref{fig:domain}).

We consider incompressible single-fluid-phase flows at low Reynolds numbers ($Re \ll 1$). The porous medium is supposed to be fully saturated with the same fluid that is present in the free-flow region. From the microscale perspective, the Stokes equations describe the fluid flow in the whole flow domain $\Omega_\mi=\Omega_\FF \cup \Omega_\ps$, which comprises the free-flow region $\Omega_\FF$ and the pore space $\Omega_\ps$  of the porous medium (Fig.~\ref{fig:domain}, left). At the macroscale, the flow system is divided into two different continuum flow regions $\Omega_\FF$ and $\Omega_\PM$ separated by the sharp fluid--porous interface $\Gamma$ (Fig.~\ref{fig:domain}, right), which is void of thermodynamic properties. In this case, two different models are applied in the two flow domains and the models need to be coupled on the interface~$\Gamma$.

\subsection{Microscale model}
\label{sec:microscale}
Under the assumptions given in \sec~\ref{sec:assumptions}, fluid flow in the whole flow domain $\Omega_\mi$ is governed by the Stokes equations 
\begin{equation} \label{eq:micro}
\deld \vec v_\mi = 0, \qquad 
- \deld \ten T(\vec v_\mi,p_\mi) - \rho \vec g = \vec 0
\qquad \text{in} \;\; \Omega_\mi,
\end{equation}
completed with the no-slip condition on the fluid--solid boundary 
\begin{equation}\label{eq:micro-no-slip}
\vec v_\mi = \vec 0 \quad \text{on} \;\; \partial \Omega_\mi \setminus \partial \Omega,
\end{equation}
and appropriate conditions on the external boundary $\partial \Omega = \Gamma_D \cup \Gamma_N$:
\begin{equation}\label{eq:micro-BC}
\vec v_\mi = \overline {\vec v} \quad \text{on} \;\; \Gamma_{D}, \qquad
\ten T(\vec v_\mi, \, p_\mi)  \vec n = \overline {\vec 
h}\hspace*{3.5mm} \text{on} \;\; \Gamma_{N}.
\end{equation}
Here, $\vec v_\mi$ and $p_\mi$ denote the fluid velocity and pressure,  $\rho$ is the fluid density, $\vec g$ is the gravitational acceleration, 
$\ten T(\vec v,p) = \mu \nabla \vec v - p \tI$ is the stress tensor, $\tI$~is the identity tensor, $\mu$ is the dynamic viscosity, $\overline {\vec v}$ and $\overline {\vec h}$ are given functions, and $\vec n$ is the unit outward normal vector on $\partial \Omega$.

\subsection{Macroscale model}
\label{sec:macroscale}

At the macroscale, we consider two different flow models in the two domains. Fluid  flow in the \textit{free-flow region} $\Omega_\FF$ is described by the Stokes equations
\begin{equation}\label{eq:FF}
\deld \vec{v}_\FF = 0, \qquad 
- \deld \ten T(\vec v_\FF,p_\FF) - \rho \vec g = \vec 0
\qquad \text{in} \;\; \Omega_\FF,
\end{equation}
where $\vec v_\FF$ is the fluid velocity and
$p_\FF$ is the pressure. 
On the external boundary of the free-flow region $\partial \Omega_\FF \setminus\Gamma = \Gamma_{D, \FF} \cup \Gamma_{N, \FF}$, we impose the following boundary conditions 
\begin{equation}\label{eq:FF-BC}
\begin{split}
\vec v_\FF = \overline {\vec v} \hspace*{3.5mm} \text{on} \;\; \Gamma_{D, \FF}, 
\qquad 
\ten T(\vec v_\FF, \, p_\FF) \vec n_\FF = \overline {\vec 
h}  \hspace*{3.5mm} \text{on} \;\; \Gamma_{N, \FF},
\end{split}
\end{equation}
where $\vec n_\FF$ is the unit outward normal vector on $\partial \Omega_\FF$. 

Fluid flow through the \textit{porous medium} $\Omega_\PM$ is described by the Darcy flow equations 
\begin{equation}\label{eq:PM}
\deld \vec v_\PM = 0, \qquad 
\vec v_\PM = -\frac{\ten K}{\mu} \lrp{\del p_\PM - \rho \vec g} \qquad \text{in} \;\; 
\Omega_\PM, 
\end{equation}
where $\vec v_\PM$ is the Darcy velocity, $p_\PM$ 
is the fluid pressure and $\ten K$ is the intrinsic permeability tensor, which is symmetric, positive definite and bounded. 
The following boundary conditions are prescribed on the external boundary of the porous-medium domain $\partial \Omega_\PM \setminus \Gamma = \Gamma_{D, \PM} \cup \Gamma_{N, \PM}$: 
\begin{equation}\label{eq:PM-BC}
\begin{split}
p_\PM = \overline p \quad \text{on} \;\; \Gamma_{D, \PM}, \qquad
\vec v_\PM \vdot \vec n_\PM = \overline v \quad \text{on} \;\; \Gamma_{N, 
\PM},
\end{split}
\end{equation}
where $\Gamma_{D, \PM} \cap \Gamma_{N, 
\PM}=\emptyset$, $\vec n_\PM$ is the unit outward normal vector on $\partial \Omega_\PM$, and $\overline p$, $\overline v$ are given functions.
Within this work, we neglect the gravitational effects setting $\vec g = \vec 0$ in equations~\eqref{eq:micro},~\eqref{eq:FF} and~\eqref{eq:PM}.

In addition to the boundary conditions on the external boundary of the coupled domain $\partial \Omega$, appropriate \textit{interface conditions} have to be set on the fluid--porous interface $\Gamma$. In this paper, we consider two different sets of coupling conditions: i)~classical conditions which are valid for parallel flows to the interface and ii)~generalized conditions which are applicable for arbitrary flow directions.

The classical set of coupling conditions which is most often used in the literature consists of the conservation of mass across the interface~\eqref{eq:IC-mass}, the balance of normal forces~\eqref{eq:IC-momentum} and the Beavers--Joseph condition \eqref{eq:IC-BJ} for the tangential velocity component 
\begin{alignat}{2}
    \vec v_\FF \vdot \vec n &= \vec v_\PM \vdot \vec n \qquad &&\text{on} 
    \;\Gamma\, , \label{eq:IC-mass} 
    \\
    \label{eq:IC-momentum}
    -\vec n \vdot \ten T\lrp{\vec v_\FF, p_\FF}\vec n &= p_\PM 
     \qquad &&\text{on} \; \Gamma \, ,
    \\
    \label{eq:IC-BJ}
    \lrp{\vec v_\FF -\vec v_\PM} \vdot \vec \tau - \frac{\sqrt{K}}{\alpha_\BJ}  
    \nabla \vec v_\FF \vec n \vdot  \vec
    \tau &= 0 \qquad &&\text{on} \; \Gamma \, .
\end{alignat}
Here, $\vec n= -\vec n_\FF  =\vec n_\PM$ is the unit vector normal to the fluid--porous interface~$\Gamma$ pointing outward from the porous-medium domain $\Omega_\PM$,  
$\vec \tau$~is the unit vector tangential to the interface (Fig.~\ref{fig:domain}, right),  $\alpha_\BJ>0$ is the Beavers--Joseph slip coefficient and $K= \vec \tau \vdot \ten K  \vec \tau \in \mathbb{R}^+$. 

The generalized interface conditions are  recently developed in~\citep{Eggenweiler_Rybak_MMS20} for arbitrary flow directions to the fluid--porous interface using homogenization and boundary layer theory. They read
\begin{alignat}{2}
    \vec v_\FF \vdot \vec n &= \vec v_\PM \vdot \vec n \qquad  &&\text{on} \;\Gamma \, , \label{eq:ER-mass}
    \\
    p_\PM &=  -\vec n \vdot \ten T\lrp{\vec v_\FF, p_\FF}\vec n - \mu N_s^{\bl}  \nabla \vec v_\FF\vec n \vdot  \vec
    \tau \qquad &&\text{on} \;\Gamma \, , \label{eq:ER-momentum}
    \\
    \vec v_\FF \vdot \vec \tau &= 
     \ell (\vec N^{\bl} \vdot \vec \tau) \nabla \vec v_\FF \vec n \vdot  \vec \tau
    - \mu^{-1}\ell^2 \sum_{j=1}^2 \left( \vec M^{j,\bl} \vdot \vec \tau   \right) \frac{\partial p_\PM}{\partial x_j}
    \qquad &&\text{on} \;\Gamma \, , \label{eq:ER-tangential}
\end{alignat}
where $N_s^{\bl} \in \mathbb{R}, \, \vec N^{\bl} = (N_1^{\bl}, N_2^{\bl})^\top \in \mathbb{R}^2$ and $\vec M^{j,\bl} = (M_1^{j,\bl}, M_2^{j,\bl})^\top  \in \mathbb{R}^{2}$ are the boundary layer coefficients with $N_1^{\bl}>0$ and $M_1^{1,\bl}>0$.

The interface condition~\eqref{eq:ER-mass} is the conservation of mass across the fluid--porous interface, coupling condition~\eqref{eq:ER-momentum} is an extension to the balance of normal forces~\eqref{eq:IC-momentum}, and condition~\eqref{eq:ER-tangential} is a generalization of the Beavers--Joseph condition~\eqref{eq:IC-BJ} on the tangential velocity component. In this paper, we reformulate condition~\eqref{eq:ER-tangential} such that it has the same analytical form as the original Beavers--Joseph interface condition~\eqref{eq:IC-BJ}:
\begin{align}
    \left( \vec v_\FF -\vec v_\PM^\text{int} \right) \vdot \vec \tau - \ell (\vec N^{\bl} \vdot \vec \tau) \nabla \vec v_\FF  \vec n \vdot  \vec \tau = 0
    \qquad \text{on} \;\Gamma \, . \label{eq:IC-BJ-mod}
\end{align}
Here, we define the interfacial velocity as
\begin{align}
    \vec v_\PM^\text{int} 
    = - \frac{\ell^2  \ten M^{\bl}}{\mu} \nabla p_\PM \, ,
\end{align}
where $\ten M^{\bl} \in \mathbb{R}^{2 \times 2}$ can be interpreted as the interfacial permeability tensor 
\begin{align*}
 \ten M^{\bl}  = \begin{pmatrix} M_1^{1,\bl} \, & M_1^{2,\bl}  \\[1.5mm] M_2^{1,\bl} \, & M_2^{2,\bl}  \end{pmatrix}
 \, .
\end{align*}

\begin{remark}
In~\citep{Sudhakar_21} alternative coupling conditions for the Sto\-kes/Darcy problem are developed. There are similarities between these conditions and the generalized interface conditions~\eqref{eq:ER-mass},~\eqref{eq:ER-momentum} and~\eqref{eq:ER-tangential} or~\eqref{eq:IC-BJ-mod}, respectively. 
The coupling condition on the tangential velocity from~\citep{Sudhakar_21} includes an interfacial porous-medium velocity instead of the Darcy velocity as the generalized condition~\eqref{eq:IC-BJ-mod}. The conservation of mass across the interface~\eqref{eq:ER-mass} is extended 
to allow transport along~$\Gamma$. The third coupling condition from~\citep{Sudhakar_21} is the generalized condition~\eqref{eq:ER-momentum} with two additional terms. One term accounts for the normal force induced at the interface due to wall-normal velocity and the other one is a higher-order term that arises during the derivation procedure.
The interface conditions developed by~\cite{Sudhakar_21} are supposed to account for arbitrary flow directions to the interface, but so far, they have only been validated for the lid driven cavity over porous bed, where the fluid flow is almost parallel to the interface.
\end{remark}

\begin{remark}
The boundary layer constants $N_s^{\bl}, \, \vec N^{\bl}$ and $\vec M^{j,\bl}$ appearing in the generalized conditions~\eqref{eq:ER-momentum} and \eqref{eq:ER-tangential} are computed based on the pore geometry and exact location of the fluid--porous interface as described in Section~\ref{sec:parameters}. 
For the particular choice of the interface position we obtain the corresponding boundary layer constants, which will change if another interface position is considered. Note that the sharp interface location can be chosen with some freedom above the solid obstacles~\citep{Eggenweiler_Rybak_MMS20}.
\end{remark}

\begin{remark}
For orthotropic porous media with $\ten K = \operatorname{diag}\{k_{11}, k_{22}\}$, we have $N_s^\bl=0$. In this case, condition~\eqref{eq:ER-momentum} reduces to the classical balance of normal forces across the fluid--porous interface given in~\eqref{eq:IC-momentum}.
\end{remark}

\begin{remark}
The coupling condition \eqref{eq:IC-BJ-mod} contains the interfacial velocity~$\vec v^\text{int}_\PM$, which is typically different from the Darcy velocity~$\vec v_\PM$ appearing in the Beavers--Joseph condition~\eqref{eq:IC-BJ}. For some porous-medium configurations it is possible to obtain $\vec v_\PM^\text{int} \vdot \vec \tau = \vec v_\PM \vdot \vec \tau$ by the appropriate choice of the fluid--porous interface location while computing the boundary layer constants~$\vec M^{j,\bl}$. 
However, this is not the case in general.
\end{remark}

\begin{remark}
For some orthotropic porous media, which satisfy the condition $\vec v_\PM^\text{int} \vdot \vec \tau = \vec v_\PM \vdot \vec \tau$ (see Remark 4), we can compute the Beavers--Joseph slip coefficient by setting 
$\alpha_\BJ = \sqrt{K}/(\ell N_1^\bl)$. 
In this case, the generalized interface  condition~\eqref{eq:IC-BJ-mod} represents the original Beavers--Joseph condition~\eqref{eq:IC-BJ}.
\end{remark}
 
\section{Computation of effective coefficients}
\label{sec:parameters}

In order to obtain computable macroscale models all effective model parameters need to be determined. For the Stokes/Darcy model~\eqref{eq:FF}--\eqref{eq:PM-BC} with the classical interface conditions~\eqref{eq:IC-mass}--\eqref{eq:IC-BJ}, these are the permeability tensor~$\ten K$ and the Beavers--Joseph slip coefficient~$\alpha_\BJ$. The permeability tensor is computed using homogenization theory~\citep{Hornung_97} based on the pore geometry. The Beavers--Joseph slip coefficient $\alpha_\BJ$ appearing in condition~\eqref{eq:IC-BJ} is an uncertain parameter, which needs to be found experimentally or fitted. This parameter depends on the exact position of the fluid--porous interface and the pore geometry in the vicinity of the interface. In this paper, we consider the interface located tangent to the top of the first row of solid inclusions as suggested in~\citep{Beavers_Joseph_67,Rybak_etal_19,Lacis_Bagheri_17}. 

Effective coefficients appearing in the Stokes/Darcy model~\eqref{eq:FF}--\eqref{eq:PM-BC} with the generalized coupling conditions~\eqref{eq:ER-mass}--\eqref{eq:ER-tangential} are the permeability $\ten K$ and the boundary layer constants $N_s^{\bl} , \, \vec N^{\bl}$ and $\vec M^{j,\bl}$. All these coefficients are computed using homogenization and boundary layer theory~\citep{Eggenweiler_Rybak_MMS20} as described in Section~\ref{sec:homogen}.

In case of fluid flows parallel to the porous bed, the Beavers--Joseph coupling condition~\eqref{eq:IC-BJ} is valid. 
For such flow regimes, we propose an efficient two-level algorithm to determine the optimal value of the Beavers--Joseph coefficient in Section~\ref{sec:two-level-alg}. Taking the optimal value $\aopt$ instead of the most commonly used $\alpha_\BJ=1$ significantly reduces the error between the pore-scale resolved and the macroscale solutions (see Section~\ref{sec:results}).

\subsection{Homogenization and boundary layer theory}\label{sec:homogen}

\begin{figure}[t]
\centerline{\includegraphics[width=0.75\textwidth]{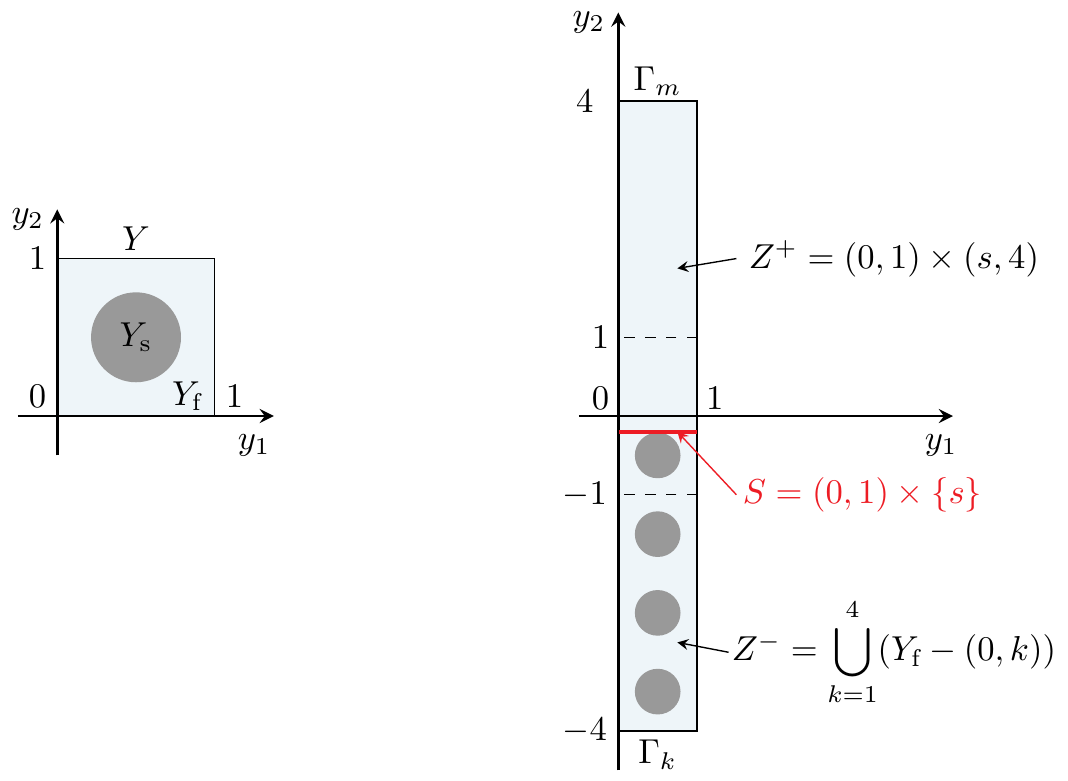}}
\caption{Unit cell $Y$ (left) and stripe $Z^\bl= \Zp \cup \Zm$ for the solution of the cell problems~\eqref{eq:cell-prob} and the boundary layer problems~\eqref{eq:BL-t-1}--\eqref{eq:BL-t-jump} and \eqref{eq:BL-t-1}--\eqref{eq:BL-t-2}, \eqref{eq:BL-t-jump-2}, respectively}
\label{fig:stripe}
\end{figure}

In order to compute the permeability tensor $\ten K= (k_{ij})_{i,j=1,2}$, we apply the theory of homogenization~\citep{Hornung_97}. This leads to 
\begin{equation}\label{eq:permeability}
k_{ij} =\ell^2 \tilde{k}_{ij}, \qquad  \tilde{k}_{ij}= \int_{Y_\text{f}} w_{i}^j \ \text{d}  \vec{y}, 
\end{equation}
where  $\vec{w}^{j} = (w_1^j,w_2^j)$ is the solution to the cell problem
\begin{equation}\label{eq:cell-prob}
\begin{split}
\nabla \vdot \; &\vec{w}^{j}  =0  \, , \quad- \Delta \vec{w}^{j}  + \nabla \pi^j = 
\vec{e}_j \quad \text{in 
$Y_\text{f}$}, \quad \int_{Y_\text{f}} \pi^j \ \text{d} \vec{y} =0 \, ,
\\
&\vec{w}^{j} =  \vec{0} \quad \text{on $\partial Y_\text{f} \setminus \partial Y$} \, ,  \quad
\{ \vec{w}^{j}, \pi^j \} \text{ is 1-periodic} \, , \quad  
\end{split}
\end{equation}
for $j=1,2$. Here, $Y_\text{f}$ denotes the fluid part of the periodicity cell $Y=(0,1)\times (0,1) = Y_\text{f} \cup Y_\text{s}$ (Fig.~\ref{fig:stripe}, left). 

The effective coefficients $N_s^{\bl}$, $\vec N^{\bl}$ and $\vec M^{j,\bl}$ appearing in conditions~\eqref{eq:ER-momentum} and~\eqref{eq:ER-tangential} are obtained from the boundary layer problems constructed in~\citep{Eggenweiler_Rybak_MMS20}. These problems are defined on the stripe $Z^\bl = \Zp \cup \Zm$, where the free-flow part $\Zp=(0,1)\times (s,4)$ and the pore space $\Zm= \displaystyle \cup_{k=1}^4 (Y_\text{f}-(0,k))$ are separated by the sharp interface $S=(0,1) \times \{s \}$ with $s < 0$. In this paper, we locate the interface $S$ directly on top of the solid inclusions (Fig.~\ref{fig:stripe}, right). 

We denote the solid boundary within the boundary layer stripe by $\partial Z^-_\text{s} = \partial Z^- \setminus (\partial Z^\bl \cup S)$, the upper and lower boundary by~$\Gamma_m$ and~$\Gamma_k$, respectively. To obtain coefficients $N_s^{\bl}$ and $\vec N^{\bl}$, we solve the boundary layer problem
\begin{align}
&\nabla \vdot \vec t^{\bl} = 0 \, , 
\ \ 
-\Delta \vec t^{\bl} + \nabla s^{\bl} = \vec 0 \quad \text{in } \Zp \cup \Zm \, , \quad 
\int_{\Gamma_k}   s^{\bl}  \, \text{d}y_1 =0  \, , \quad
\label{eq:BL-t-1}
\\ 
&
\frac{\partial t_1^{\bl}}{\partial y_2} = t_2^{\bl} =  0 \quad \text{on } \Gamma_m  \, , \quad
\vec t^{\bl} = \vec 0 \quad \text{on } 
\partial \Zm_\text{s}\, , 
\quad 
\{\vec t^{\bl}, s^{\bl}\} \text{ is $y_1$-periodic} \, ,
\label{eq:BL-t-2}
\\[1.5mm] 
&
\big\llbracket ( \nabla \vec t^{\bl} - s^{\bl} \ten{I} ) \vec{e}_2\big\rrbracket_{S} =  -\vec{e}_1 \, , \ \
\big\llbracket \vec t^{\bl} \big\rrbracket_{S} = \vec 0 \quad \text{on } S \, , \quad \ \vec t^{\bl}  = \vec 0 \quad \text{on } \Gamma_k \, , 
\label{eq:BL-t-jump}
\end{align}
and integrate the solutions afterwards 
\begin{align}
\vec N^{\bl}=\left(\int_S  t_1^{\bl} \ \text{d} y_1, 0 \right) \, ,  
\qquad
N_s^\bl = \int_S s^{\bl} \ \text{d} y_1\, .
\end{align}
The boundary layer constants $\vec M^{j,\bl}$ are obtained by solving two flow problems~\eqref{eq:BL-t-1}--\eqref{eq:BL-t-2} with the following jump conditions across the interface $S$: 
\begin{align}
\big\llbracket ( \nabla \vec t^{\bl} - s^{\bl} \ten{I} ) \vec{e}_2\big\rrbracket_{S} =  \left( \nabla \vec w^{j} - \pi^j \right) \vec{e}_2 \, , \quad
\big\llbracket \vec t^{\bl} \big\rrbracket_{S} = \vec w^j -\tilde k_{2j}\vec{e}_2  \ \, \text{on } S \, ,
\label{eq:BL-t-jump-2}
\end{align}
and integrating their solutions
\begin{align}
\vec M^{j,\bl}=\left(\int_S t_1^{\bl} \ \text{d} y_1, 0 \right) \, ,  \quad j=1,2 \, .
\end{align}

\subsection{Two-level numerical algorithm}
\label{sec:two-level-alg}

In this section, we propose an efficient two-level numerical algorithm to determine the optimal value of the Beavers--Joseph slip coefficient $\alpha_\BJ$ for coupled flow problems, where the Beavers--Joseph condition~\eqref{eq:IC-BJ} is valid.  
The algorithm is based on the minimization of the difference between the pore-scale resolved and the macroscale numerical simulation results. 
This difference is quantified by the errors $\epsilon_{f, \, c, \, \alpha}$ defined along the fixed cross section $x_1=c$ as follows
\begin{equation}
    \begin{aligned}
    \epsilon_{f, \, c, \, \alpha} = \frac{\|f_{\ma, \, \alpha}(c, \cdot)-f_\mi(c, \cdot)\|}{\|f_\mi(c, \cdot)\|}, \qquad \|f(c, \cdot)\|^2=\int_h^H f(c,x_2)^2 \, \mathrm{d}x_2 \,,
    \end{aligned}
    \label{eq:min}
\end{equation}
where $f_\mi$ is the solution of the pore-scale resolved problem~\eqref{eq:micro}--\eqref{eq:micro-BC}  and $f_{\ma, \, \alpha}$ is the solution of the macroscale problem \eqref{eq:FF}--\eqref{eq:IC-BJ} with $\alpha_\BJ = \alpha$. In case the generalized conditions \eqref{eq:ER-mass}--\eqref{eq:ER-tangential} are applied, we consider $f_{\ma}$ for the macroscale solution and denote the error by $\epsilon_{f, \, c}$. 

The Clough--Tocher interpolation is applied to handle the two different meshes used for numerical simulations of pore-scale resolved and macroscale models. Due to the presence of solid obstacles at the microscale, an adaptive triangular mesh is applied in the whole fluid domain $\Omega_\mi$ for the pore-scale resolved simulations. The macroscale solutions are computed on the staggered Cartesian mesh in the coupled domain $\Omega$. To compare the microscale and macroscale solutions we interpolate the pore-scale simulation results using the Clough--Tocher interpolation~\citep{Alfeld_84,Nielson_83}, which provides $C^1$-functions and operates on triangulated data.

Instead of solving the Stokes/Darcy problem~\eqref{eq:FF}--\eqref{eq:IC-BJ} for a huge range of parameters $\alpha_\BJ$ to determine the one yielding the minimal error \eqref{eq:min}, we propose the following efficient two-level procedure. In the first step (Level~1) of the two-level algorithm we compute a coarse approximation of the optimal value of the Beavers--Joseph parameter. This value is determined in the second refinement step (Level~2).
\\
\emph{\underline{Level 1}:} We perform macroscale numerical simulations for the Stokes/Darcy problem~\eqref{eq:FF}--\eqref{eq:IC-BJ} considering the following range of the Beavers--Joseph slip coefficient $\mathcal{A}_\BJ = \{0.01, 1, 2,\ldots, M\}$ with $M \in \mathbb{N}$, and compute the corresponding relative errors~\eqref{eq:min}. We consider the cubic spline interpolation of the $(M+1)$ error values and apply simulated annealing (\algo~\ref{Algorithm1}) to find a coarse approximation~$\alpha_\BJ^*$ of the optimal Beavers--Joseph parameter. This value is not accurate enough due to interpolation, therefore, the refinement step (Level~2) is applied. We set the lower boundary $\alpha_\BJ = 0.01$ due to positivity of the Beavers--Joseph parameter $\alpha_\BJ>0$.

\begin{algorithm}[H]
	\caption{Simulated annealing}
 	\begin{algorithmic}[1]
 	    \Require function $f(x)$, interval $[x_l, \, x_r]$
 	    \Ensure $x_{\text{opt}}$ = optimal $x$
 		\State Choose $x_0 \in [x_l, \, x_r]$ and set $A_0 = 1$, \, $c_{\text{grow}}>1$, \,   $c_{\text{shrink}}<1$
 		\State Choose a monotonically decreasing null sequence $(T_n)_{n \in \mathbb{N}_0}$
		\State $x_{\text{approx}} = x_0$, \, $x_{\text{opt}}= x_0$, \, $n = 0$
 		\Repeat
 		\State Choose $r_n \sim \mathcal{N}(0,1)$, \,
 		$x_{n+1} = x_{n}+A_n\, r_n$
 		\If {$f(x_{n+1}) \leq f(x_{\text{approx}})$}
 		\State $x_{\text{approx}} = x_{{n}+1}$, \, $A_{n+1} = A_n \, c_{\text{grow}}$
 		\ElsIf{$\operatorname{rand}([0,1])< \operatorname{exp}{\left(\left(f\left(x_{n+1}\right)-f\left(x_{\text{opt}}\right)\right)/ T_n\right)}$}
		\State $x_{\text{approx}} = x_{n+1}$, \, $A_{n+1} = A_n \, c_{\text{grow}}$
 		\Else
 		\State $A_{n+1} = A_n \, c_{\text{shrink}}$
 		\EndIf
 		\If{$f(x_{\text{approx}})<f(x_{\text{opt}})$}
 		\State $x_{\text{opt}} = x_{\text{approx}}$
 		\EndIf
 		\State $n=n+1$
 		\Until{$T_n < \varepsilon_\text{tol}$ or $n < N_\text{max}$}
 	\end{algorithmic}
 	\label{Algorithm1}
 \end{algorithm}

\noindent
In~\algo~\ref{Algorithm1}, we choose the following monotonically decreasing null sequence $T_n = a\operatorname{exp}(-n/b)$ with $a=0.005$ and $b=100$. We set $c_{\text{grow}} = 1.1$, $c_{\text{shrink}}=0.99$, $\varepsilon_\text{tol} = 10^{-12}$ and $N_\text{max}=10^5$. \\

\noindent
\emph{\underline{Level 2}:} If the value of the Beavers--Joseph parameter  determined in Level~1 is 
$\alpha_\BJ^* \in [0.6,M]$, we round $\alpha_\BJ^*$ to the first digit after comma,
set \mbox{$m=10$} and consider the following candidates for the optimal slip coefficient $\operatorname{round}(\alpha_\BJ^*,1) + 0.5 - j/m$. 
Otherwise, we consider the candidates $0.01$ and $\lceil \alpha_\BJ^*\rceil  - j/m$ for $j= 0, \ldots, m-1$. 
The optimal value $\alpha_\BJ^{\text{opt}}$ of the Beavers--Joseph parameter is the candidate which yields the smallest relative error~\eqref{eq:min}.
Since the error is changing only slightly when considering more candidates ($m>10$) for the numerical simulation results presented in \sec~\ref{sec:results}, it is sufficient to determine $\alpha_\BJ^{\text{opt}}$ up to the first digit after comma.

  \begin{algorithm}[H]
     \caption{Two-level algorithm}
     \begin{algorithmic}[1]
       \Require cross section $c\in [0,L]$ and $m$, $M$
       \Ensure $\alpha_\BJ^{\text{opt}}$ = optimal $\alpha_\BJ$
         \State \textbf{------------ Level 1 ------------}
         \For{$i=0$ \TO $M$}
         \State Solve Stokes/Darcy problem \eqref{eq:FF}--\eqref{eq:IC-BJ} with $\alpha_\BJ \in \mathcal{A}_\BJ$
         \State Compute errors $\epsilon_{f,\,c,\, \alpha_\BJ}$ using equation \eqref{eq:min} 
         \EndFor
         \State $f_{\mathrm{int}}$ = interpolate $\epsilon_{f,\,c,\, \alpha_\BJ}$ using cubic splines
         \State $\alpha_{\BJ}^*$ = \algo~\ref{Algorithm1}$\left(f_{\mathrm{int}},\,[0.01,M]\right)$
     \State \textbf{------------ Level 2 ------------}
     \If{$\alpha_{\BJ}^* \geq 0.6$}
         \State $\alpha_\BJ^* =\operatorname{round}(\alpha_\BJ^*,1) +0.5$
     \Else
         \State $\alpha_{\BJ}^* = \lceil\alpha_{\BJ}^*\rceil =1$
     \EndIf
     \State $\epsilon_{\min} = \infty$
     \For{$j=0$ \TO $m$}
         \If{$j==m$ \textbf{and} $\alpha_{\BJ}^* == 1$}
            \State $\alpha_m = 0.01$
         \Else
            \State $\alpha_j = \alpha_{\BJ}^* -j/m$ 
         \EndIf
     \State Solve Stokes/Darcy problem \eqref{eq:FF}--\eqref{eq:IC-BJ} with $\alpha_\BJ = \alpha_j$
     \State Compute errors $\epsilon_{f,\,c,\, \alpha_j}$ using equation \eqref{eq:min}
     \If{$\epsilon_{f,\,c,\,  \alpha_j} < \epsilon_{\min}$}
         \State $\epsilon_{\text{min}} = \epsilon_{f,\,c,\, \alpha_j}$, \, $\alpha_{\BJ}^{\text{opt}} = \alpha_j$
     \EndIf
     \EndFor
     \end{algorithmic}
     \label{Algorithm2}
 \end{algorithm}

\begin{remark}
The upper boundary $M$ for the Beavers--Joseph coefficient $\alpha_\BJ$ should be chosen depending on the application of interest. The number of candidates $m$ in Level 2 can be increased in order to determine $\aopt$ more accurately, if necessary. Alternative interpolation techniques can be considered in Level 1.
\end{remark}

\section{Numerical results}
\label{sec:results}

In this section, we study the applicability of the classical interface conditions~\eqref{eq:IC-mass}--\eqref{eq:IC-BJ} and the generalized interface conditions \eqref{eq:ER-mass}--\eqref{eq:ER-tangential}, where  condition \eqref{eq:ER-tangential} is formulated in 
the same form \eqref{eq:IC-BJ-mod} as the original Beavers--Joseph condition \eqref{eq:IC-BJ}, to describe fluid flows in coupled free-flow and porous-medium systems.
We consider two flow problems: i)~pressure-driven flow, where fluid flow is parallel to the porous layer, and ii)~general filtration, where flow is arbitrary to the fluid--porous interface. 
We show that one can find an optimal value of the Beavers--Joseph slip coefficient $\alpha_\BJ$ appearing in condition~\eqref{eq:IC-BJ} for unidirectional flows parallel to the interface. 
We determine this value using the efficient numerical algorithm proposed in~Section~\ref{sec:two-level-alg}.
Alternatively, one can also apply the generalized coupling conditions~\eqref{eq:ER-mass},~\eqref{eq:ER-momentum} and~\eqref{eq:IC-BJ-mod}, where all model parameters are computed based on the pore geometry using homogenization and boundary layer theory (see Section~\ref{sec:homogen}).
We demonstrate that for arbitrary flows to the porous layer, it is not possible to find an optimal value of the Beavers--Joseph coefficient. In this case the coupling condition
~\eqref{eq:IC-BJ-mod} should be used instead of the classical 
Beavers--Joseph condition \eqref{eq:IC-BJ}.

We study various porous-medium geometrical configurations considering different shapes of solid inclusions (circular, rectangular, elliptical) and different arrangements of solid grains (in-line, staggered) leading to isotropic, orthotropic and anisotropic media. We consider different porous-medium geometries, which lead to the same permeability and similar porosity (Tab.~\ref{tab:geoms}, geometries $G_1$ and $G_3$); porous media containing the same type of solid inclusions of different sizes which result in different interface roughness, permeability and porosity (Tab.~\ref{tab:geoms}, geometries $G_1$, $G_2$ and $G_4$); and pore geometries having the same porosity and the same interface roughness but different permeability tensors (Tab.~\ref{tab:geoms}, geometries $G_1$, $G_4$ and $G_5$, $G_6$). All considered porous-medium geometries are periodic such that homogenization theory is applied to compute the permeability tensor and the boundary layer constants as described in~Section~\ref{sec:homogen}. We set the dynamic viscosity of the fluid $\mu=1$ for all flow problems.

\begin{table}[!h]
\caption{Permeability values $k_{ij}$, porosity $\phi$ and boundary layer constants for different pore geometries}
    \begin{figure}[H]
    \includegraphics[scale=0.7]{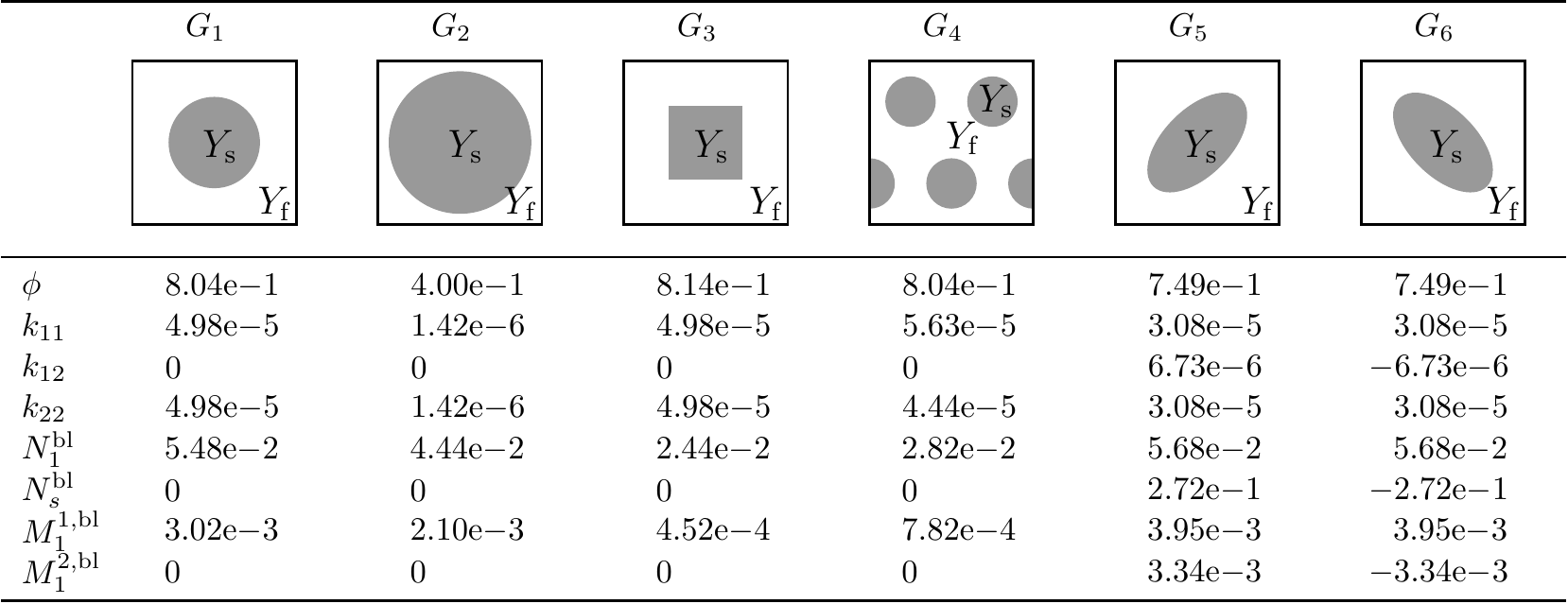}
    \end{figure}
    \label{tab:geoms}
\end{table}

The microscale problem \eqref{eq:micro}--\eqref{eq:micro-BC}, the unit cell Stokes problems~\eqref{eq:cell-prob} and the boundary layer problems \eqref{eq:BL-t-1}--\eqref{eq:BL-t-jump} and \eqref{eq:BL-t-1}, \eqref{eq:BL-t-2},  \eqref{eq:BL-t-jump-2} are solved using the software package \textsc{FreeFEM++} with the Taylor--Hood finite elements~\citep{Hecht_12}. 
The macroscale Stokes/Darcy problem~\eqref{eq:FF}--\eqref{eq:IC-BJ} completed by the classical interface conditions~\eqref{eq:FF}--\eqref{eq:PM-BC} or the generalized conditions~\eqref{eq:ER-mass}, \eqref{eq:ER-momentum}, \eqref{eq:IC-BJ-mod} is discretized using the finite volume method on staggered Cartesian grids  conforming at the fluid--porous interface with the grid size $h_1=h_2=1/800$. The optimal value of the Beavers--Joseph slip coefficient $\aopt$ is computed using the proposed two-level numerical algorithm (Alg.~\ref{Algorithm2}).


\subsection{Pressure-driven flow}
\label{sec:pressure-driven}

In this section, we study a flow scenario where the flow is parallel to the fluid--porous interface (pressure-driven flow).
We consider the free-flow domain $\Omega_\FF =[0,1]\times [0,0.5]$, the porous medium $\Omega_\PM=[0,1]\times [-0.5,0]$ and the sharp fluid--porous interface $\Gamma=[0,1]\times \{0\}$. 
We analyze different porous-medium geometries (Tab.~\ref{tab:geoms})
and study the influence of the microscale interface roughness, permeability $\ten K$ and porosity $\phi$ 
on the Beavers--Joseph slip coefficient.

To describe a pressure-driven flow, we impose the following boundary conditions for the microscale model \eqref{eq:micro}--\eqref{eq:micro-BC}:
\begin{equation}\label{eq:parallel-BC-micro}
    \overline{\vec{v}} = \vec 0 \text{\; on \; }\Gamma_{D},\qquad
    \overline{\vec{h}} = (0,-31.75) \text{\; on \; }\Gamma_{N}^{\mathrm{in}},\qquad 
    \overline{\vec{h}} = \vec 0 \text{\; on \; }\Gamma_{N}^{\mathrm{out}},\\
\end{equation}
where  $\Gamma_{N}^{\mathrm{in}} = \{0\} \times [-0.5,0.5]$, $\Gamma_{N}^{\mathrm{out}}= \{ 1\} \times [-0.5,0.5]$, $\Gamma_{D} = \partial \Omega \setminus (\Gamma_{N}^\mathrm{in} \cup \Gamma_{N}^\mathrm{out})$. 
For the coupled macroscale model \eqref{eq:FF}--\eqref{eq:IC-BJ}, we set
\begin{equation}
\begin{aligned}
    \overline{\vec{v}} &=  \vec 0 &&\text{\; on \; }\Gamma_{D,\FF},
    \qquad \quad\overline v = 0 &&\text{\; on \; }\Gamma_{N,\PM}, 
    \\ 
    \overline{\vec{h}} &= (0,- 31.75) &&\text{\; on \; }\Gamma_{N,\FF}^{\mathrm{in}},
    \qquad \quad \overline p = 31.75 &&\text{\; on \; }\Gamma_{D,\PM}^\mathrm{in}, 
    \\
    \overline{\vec{h}} &= \vec 0 &&\text{\; on \; }\Gamma_{N,\FF}^{\mathrm{out}}, \qquad \quad \overline p = 0 &&\text{\; on \; }\Gamma_{D,\PM}^\mathrm{out},
\end{aligned}\label{eq:parallel-BC-macro}
\end{equation}
where $\Gamma_{D,\FF}= \Gamma_{D}\cap \partial \Omega_\FF$, $\Gamma_{N,\PM}= \Gamma_{D}\cap \partial \Omega_\PM$,  $\Gamma_{N, \FF}^\mathrm{in/out}= \Gamma_{N}^\mathrm{in/out} \cap \partial \Omega_\FF$, and $\Gamma_{D,\PM}^\mathrm{in/out}= \Gamma_{N}^\mathrm{in/out} \cap \partial \Omega_\PM$.
Boundary conditions~\eqref{eq:parallel-BC-micro} and~\eqref{eq:parallel-BC-macro} lead to unidirectional flow parallel to the fluid--porous interface, where the normal component of velocity is zero and the pressure field is linear. 
For this flow problem, the macroscale pressure and the normal velocity component do not dependent on the choice of the Beavers--Joseph parameter. Therefore, we provide velocity profiles only for the tangential velocity component in the middle of the domain at $x_1=0.5$. 

Below we study different pore geometries. To compute the permeability for geometry $G_1$, the unit cell problems~\eqref{eq:cell-prob} are solved using an adaptive mesh, where the fluid part $Y_{\text{f}}$ is partitioned into approximately 35~000 elements. For solving the pore-scale problem~\eqref{eq:micro}--\eqref{eq:micro-BC} in the entire flow domain~$\Omega_\mi$, an adaptive mesh with approximately 330~000 elements is used. We set the upper boundary $M=10$ for the Beavers--Joseph slip coefficient in the proposed two-level algorithm (\algo~\ref{Algorithm2}) in order to consider a broader range of values in comparison to~\cite{Beavers_Joseph_67}.

\subsubsection{Geometry $G_1$}
\label{sec:val-1-eps20}

\begin{figure}[H]
\includegraphics[scale=0.16]{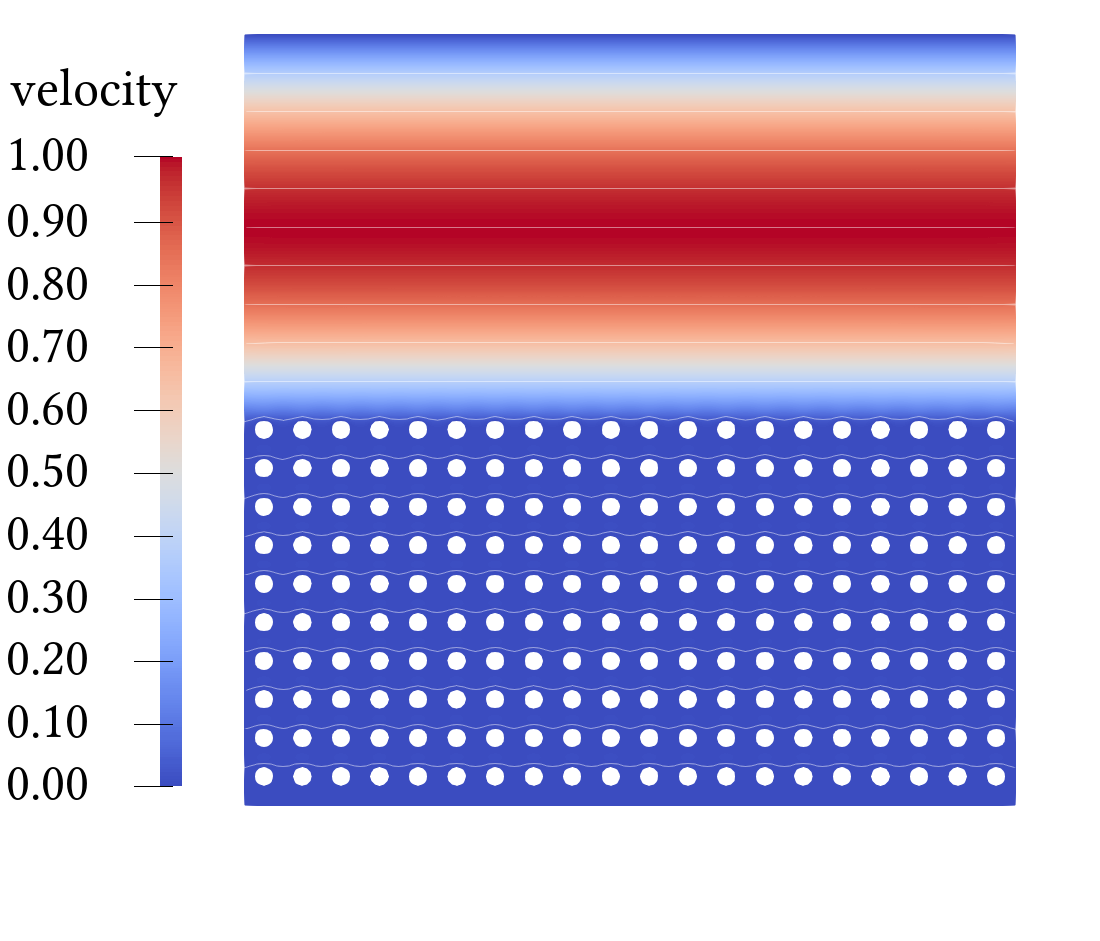} 
\includegraphics[scale=0.66]{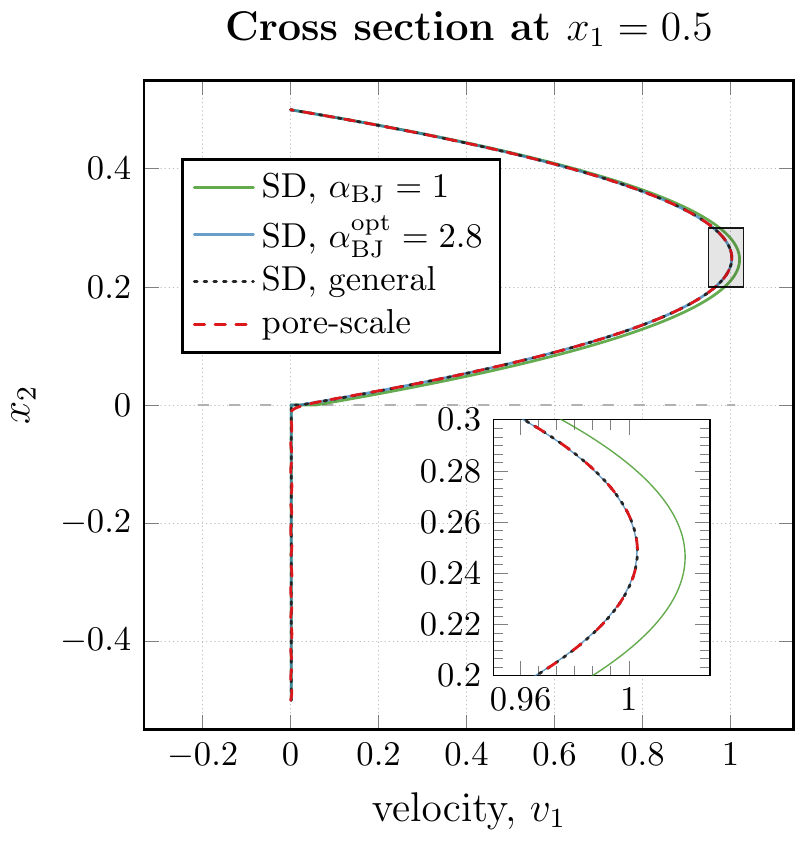}
\caption{Microscale velocity field (left) and tangential velocity profiles (right) for the pressure-driven flow and geometry $G_1$}
\label{fig:val-1-basis-g1}
\end{figure}

In this case, the porous medium is constructed by $20 \times 10$ periodically distributed circular solid inclusions presented in~Tab.~\ref{tab:geoms}, which are arranged in line (Fig.~\ref{fig:val-1-basis-g1}, left). This yields the characteristic pore size $\ell=1/20$ and the radius of inclusions is $r=0.25\ell$. The described pore-scale geometrical configuration leads to a highly porous ($\phi = 0.804$) isotropic medium with the permeability tensor $\ten K = \operatorname{diag}\{k_{11}, k_{22}\}$, $k_{11}=k_{22}$ (Tab.~\ref{tab:geoms}). The values of the boundary layer constants appearing in the generalized coupling conditions~\eqref{eq:ER-mass}--\eqref{eq:ER-tangential} are also provided in Tab.~\ref{tab:geoms}.

The pore-scale velocity field is presented in~Fig.~\ref{fig:val-1-basis-g1} (left). The optimal value of the Beavers--Joseph slip coefficient is $\alpha_\BJ^\text{opt} = 2.8$. In~Fig.~\ref{fig:val-1-basis-g1} (right), we compare two macroscale Stokes/Darcy problems~\eqref{eq:FF}--\eqref{eq:IC-BJ} with the standard value of the Beavers--Joseph parameter $\alpha_\BJ=1$ (profile: SD,~$\alpha_\BJ=1$) and the optimal value $\aopt = 2.8$ (profile: SD,~$\aopt=2.8$) against the pore-scale resolved simulations (profile: pore-scale). In addition, we provide simulation results corresponding to the Stokes/Darcy problem~\eqref{eq:FF}--\eqref{eq:PM-BC} with the generalized interface conditions~\eqref{eq:ER-mass}--\eqref{eq:ER-tangential} (profile: SD, general). The corresponding errors between the pore-scale and the macroscale simulation results are presented in~Tab.~\ref{tab:alpha-values}. The macroscale simulation results for the classical interface conditions with $\aopt =2.8$ and  the generalized conditions agree significantly better to the pore-scale resolved results than  the classical conditions with $\alpha_\BJ =1$.

The optimal value of the Beavers--Joseph slip coefficient $\aopt$ is independent of velocity magnitude. Several variants of $\overline{\vec h}$ and $\overline p$ in the boundary conditions \eqref{eq:parallel-BC-micro} and \eqref{eq:parallel-BC-macro} have been analyzed, all leading to the same optimal parameter $\aopt = 2.8$.

\subsubsection{Geometry $G_2$}
\label{sec:g1}

\begin{figure}[H]
\includegraphics[scale=0.16]{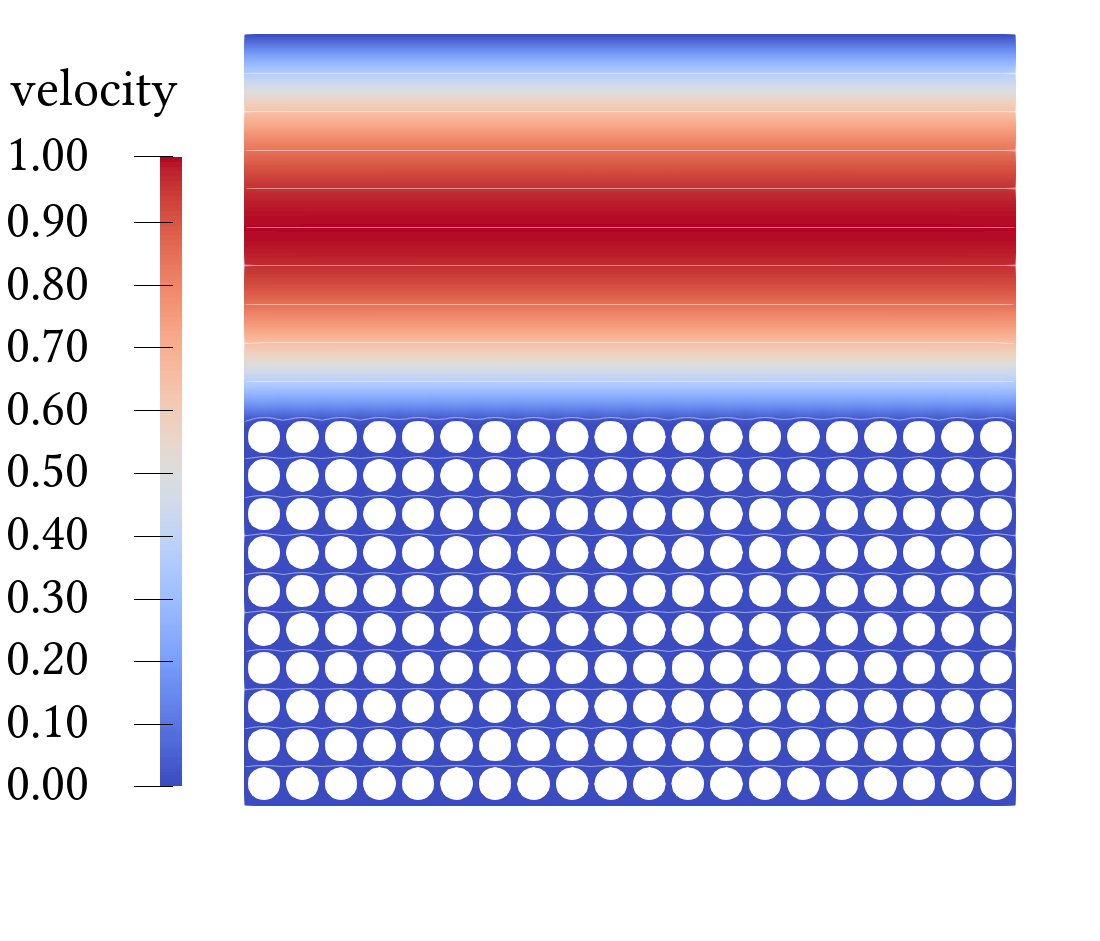} 
\includegraphics[scale=0.66]{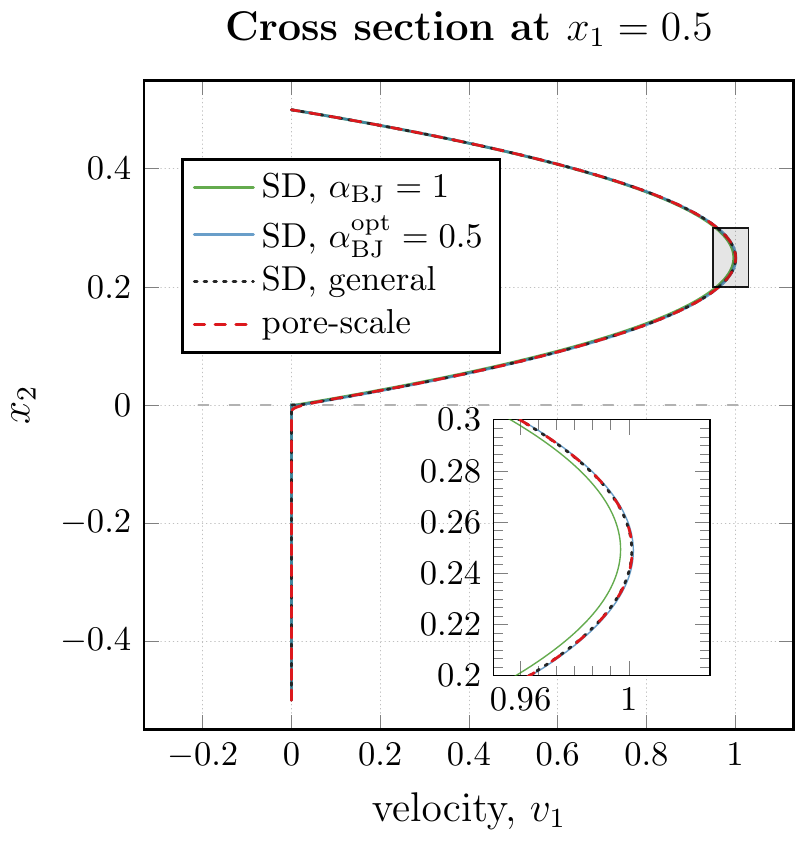}
\caption{Microscale velocity field (left) and tangential velocity profiles (right) for the pressure-driven flow and geometry $G_2$}
\label{fig:val-1-g2}
\end{figure}

In order to obtain a porous medium with moderate porosity $\phi = 0.4$, we increase the radius of solid inclusions in comparison to geometry~$G_1$. In this case, we consider $20\times 10$ circular solid inclusions ($\ell = 1/20$) with radius $r=\ell\sqrt{(1-\phi)/\pi} \approx 0.437 \ell$ arranged in line (Fig.~\ref{fig:val-1-g2}, left). 
This geometrical configuration leads again to an isotropic porous medium and the permeability values together with the boundary layer coefficients are presented in~Tab.~\ref{tab:geoms}. 
For geometry~$G_2$, the optimal value of the Beavers--Joseph parameter is $\aopt=0.5$ (Tab.~\ref{tab:alpha-values}).

In Fig.~\ref{fig:val-1-g2}, we provide velocity profiles corresponding to the pore-scale model, the Stokes/Darcy model with the classical interface conditions~\eqref{eq:IC-mass}--\eqref{eq:IC-BJ} for the two values of the Beavers--Joseph parameter~$\alpha_\BJ=1$ and $\aopt=0.5$, and the Stokes/Darcy model with the generalized coupling conditions~ \eqref{eq:ER-mass}--\eqref{eq:ER-tangential}. The corresponding errors between the pore-scale resolved and macroscale solutions are presented in Tab.~\ref{tab:alpha-values}. The classical interface conditions with the optimal value $\aopt=0.5$ as well as the generalized conditions provide more accurate results than the classical conditions with the traditionally used value $\alpha_\BJ=1$.

\subsubsection{Geometry $G_3$} 

Here, we study the dependency of the Beavers--Joseph parameter on 
the 
pore-scale interface roughness. Therefore, we construct a porous medium which has the same permeability and similar porosity as geometry $G_1$ but different shape of solid grains. We consider a porous medium consisting of $20 \times 10$ in-line arranged squared solid inclusions ($\ell = 1/20$) of length $a=0.2154\,\ell$ yielding an isotropic medium (Tab.~\ref{tab:geoms}).

\begin{figure}[H]
\includegraphics[scale=0.16]{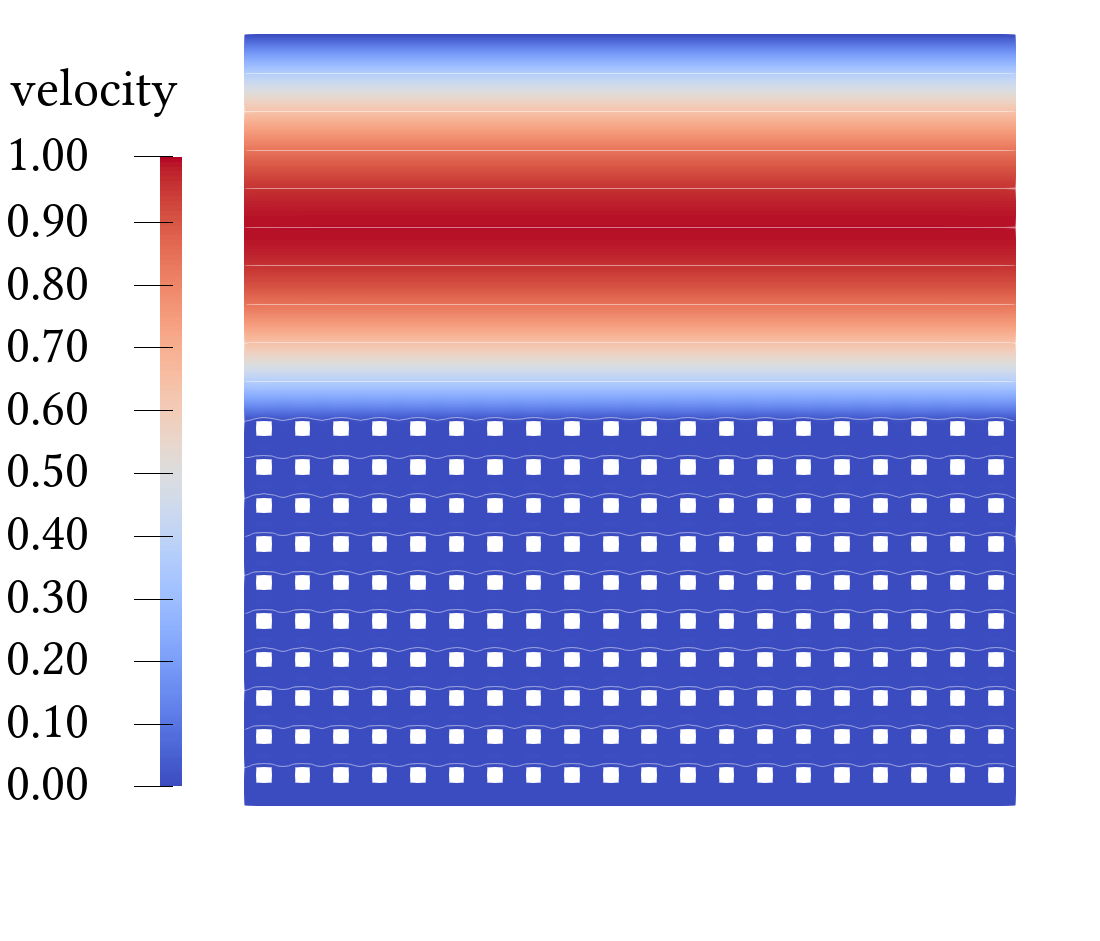} 
\includegraphics[scale=0.66]{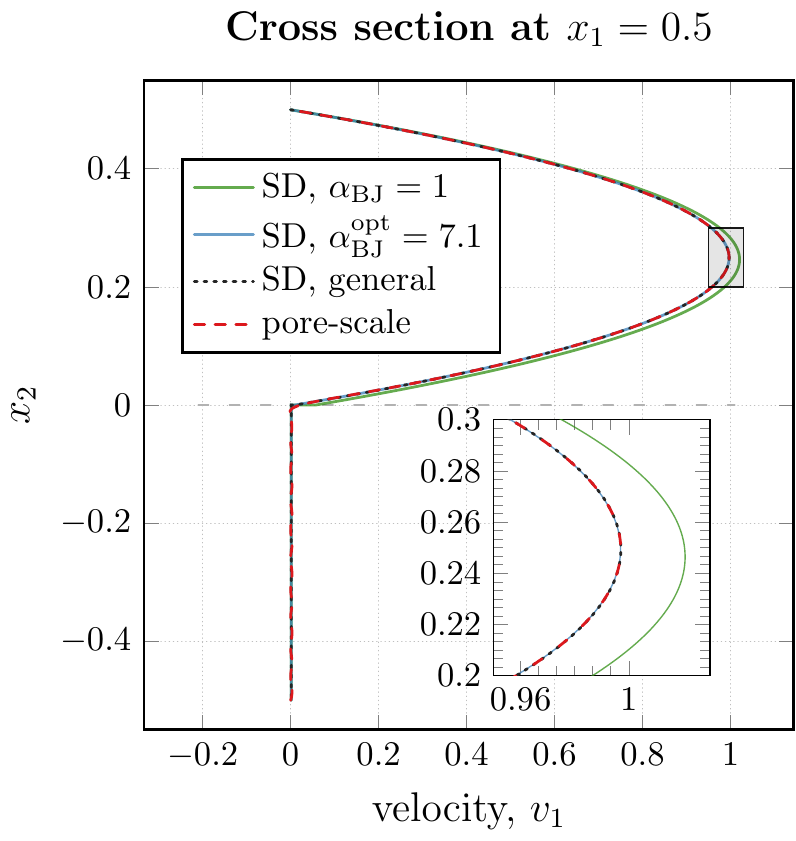}
\caption{Microscale velocity field (left) and tangential velocity profiles (right) for the pressure-driven flow and geometry $G_3$}
\label{fig:val-1-g3}
\end{figure}

We provide the pore-scale velocity field and the profiles of the tangential velocity component in the middle of the domain in~Fig.~\ref{fig:val-1-g3}. The optimal value of the Beavers--Joseph parameter for this configuration is found by Alg.~\ref{Algorithm2} to be $\alpha_\BJ^\text{opt}=7.1$ (Tab.~\ref{tab:alpha-values}). Comparing this result with the one for geometry~$G_1$ ($\alpha_\BJ^\text{opt} = 2.8$) having the same permeability and almost the same porosity, but different pore morphology, we conclude that $\alpha_\BJ$ depends not only on the effective properties of the medium but also on the pore geometry including microscale interface roughness. The modified interface condition~\eqref{eq:IC-BJ-mod} and the Beavers--Joseph condition~\eqref{eq:IC-BJ} with the optimal value $\alpha_\BJ^\text{opt} = 7.1$ provide significantly better results than the Beavers--Joseph condition with $\alpha_\BJ=1$.

\subsubsection{Geometry $G_4$}
\label{sec:val-1-staggered}
\begin{figure}[t]
    \label{fig:pd-G4}
\includegraphics[scale=0.16]{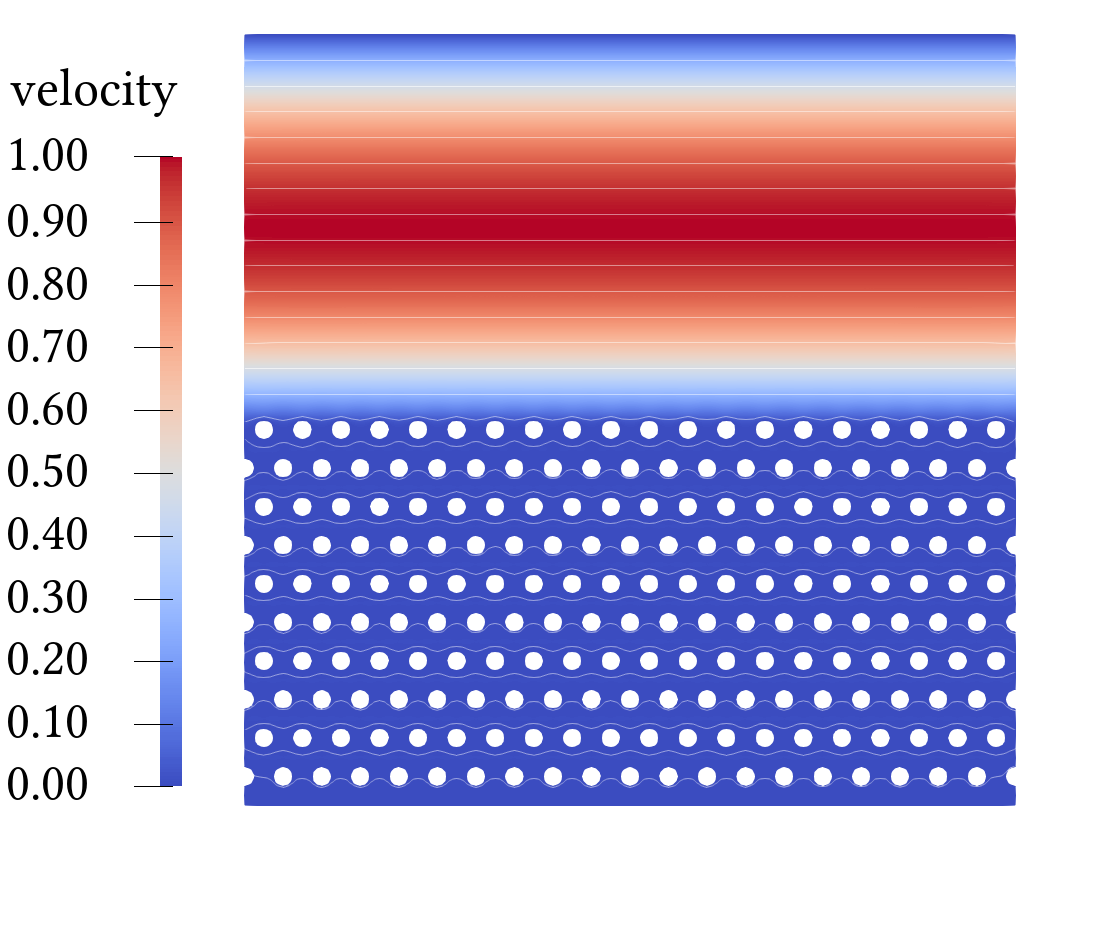} 
\includegraphics[scale=0.66]{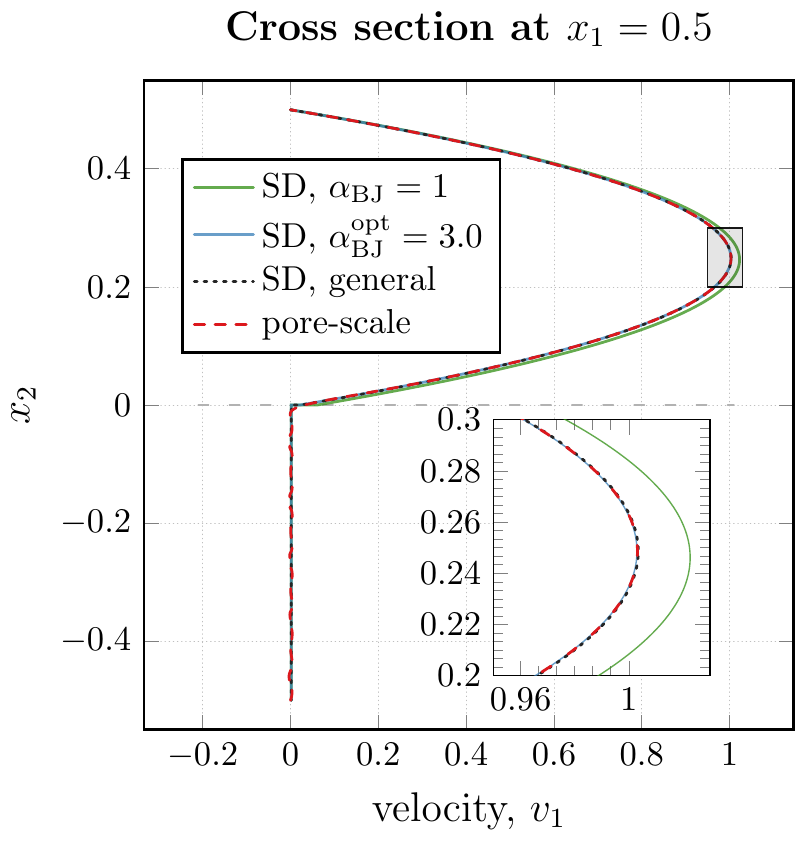}
\caption{Microscale velocity field (left) and tangential velocity profiles (right) for the pressure-driven flow and geometry $G_4$}
\label{fig:val-1-g4}
\end{figure}

In this test case, we construct a porous medium having the same interface roughness and the same porosity 
as geometry $G_1$, but different permeability. For this purpose, we consider $20 \times 10$ circular solid inclusions arranged in a staggered manner (Tab.~\ref{tab:geoms},~Fig.~\ref{fig:val-1-g4}). This leads to $\ell=1/10$ and an orthotropic porous medium with $\ten K = \operatorname{diag}\{k_{11}, k_{22}\}$,  $k_{11}\neq k_{22}$. The radius of the circular solid grains is $r=0.125\, \ell$. The permeability values and the boundary layer constants in the generalized interface conditions~\eqref{eq:ER-mass}--\eqref{eq:ER-tangential} are given in Tab.~\ref{tab:geoms}. 

For orthotropic porous media, there exist different approaches in the literature, e.g.,~\citep{Discacciati_Miglio_Quarteroni_02,Layton_Schieweck_Yotov_03, Eggenweiler_Rybak_20} to compute $\sqrt{K}$ appearing in the interface condition~\eqref{eq:IC-BJ}. As already mentioned in~Section~\ref{sec:macroscale}, the first one  is to take $\sqrt{K} = \sqrt{\vec \tau \vdot \ten K  \vec \tau}$, which leads to $\sqrt{K} = \sqrt{k_{11}}$ for our setting. The second interpretation is $\sqrt{K} = \sqrt{\operatorname{tr}(\ten K) /d }$, where $d$ is the number of space dimensions. In our case, we obtain~$\sqrt{K} = \sqrt{{(k_{11}+k_{22})}/{2}}$. The optimal values of the slip coefficient~$\alpha_\BJ$ for these two cases are presented in~Tab.~\ref{tab:alpha-values}. 
Note that taking the second approach for the computation of $\sqrt{K}$ we obtain the same value $\aopt = 2.8$ as for geometry~$G_1$, where the microscale surface roughness is exactly the same. 

The choice of $\sqrt{K}$ influences the optimal Beavers--Joseph parameter~$\aopt$, however, the ratio $\sqrt{K}/\aopt$ is almost the same for both interpretations of $\sqrt{K}$ with the corresponding value~$\aopt$.
Therefore, in Fig.~\ref{fig:val-1-g4}~(right) we present only the macroscale simulation results for the first approach $\sqrt{K} = \sqrt{k_{11}}$. The microscale velocity field corresponding to geometry~$G_4$ is provided in~Fig.~\ref{fig:val-1-g4} (left).
We observe that the macroscale simulation results for the classical interface conditions with $\aopt =3.0$ as well as for the generalized conditions agree very well to the pore-scale results, however, this is not the case for the classical conditions with $\alpha_\BJ =1$.

\subsubsection{Geometries  $G_5$ and $G_6$}
Now we consider two anisotropic porous media with full permeability tensors~$\ten K$. The porous media are composed by $20\times 10$ elliptical solid inclusions arranged in line ($\ell=1/20$) tilted to the right (Tab.~\ref{tab:geoms}, geometry $G_5$) and tilted to the left (Tab.~\ref{tab:geoms}, geometry $G_6$). The semi-axes are $a=0.4 \ell$ and $b=0.2 \ell$ and the ellipses are rotated clockwise and  counter-clockwise by $45^\circ$, respectively. Geometries $G_5$ and $G_6$ lead to different permeability tensors (Tab.~\ref{tab:geoms}), however, the value of $\sqrt{K} = \sqrt{\vec \tau \vdot \ten K  \vec \tau}$ in the Beavers--Joseph condition~\eqref{eq:IC-BJ} is the same for our setting. 
Moreover, both geometries provide the same interface roughness. 

\begin{figure}[!ht]
\includegraphics[scale=0.16]{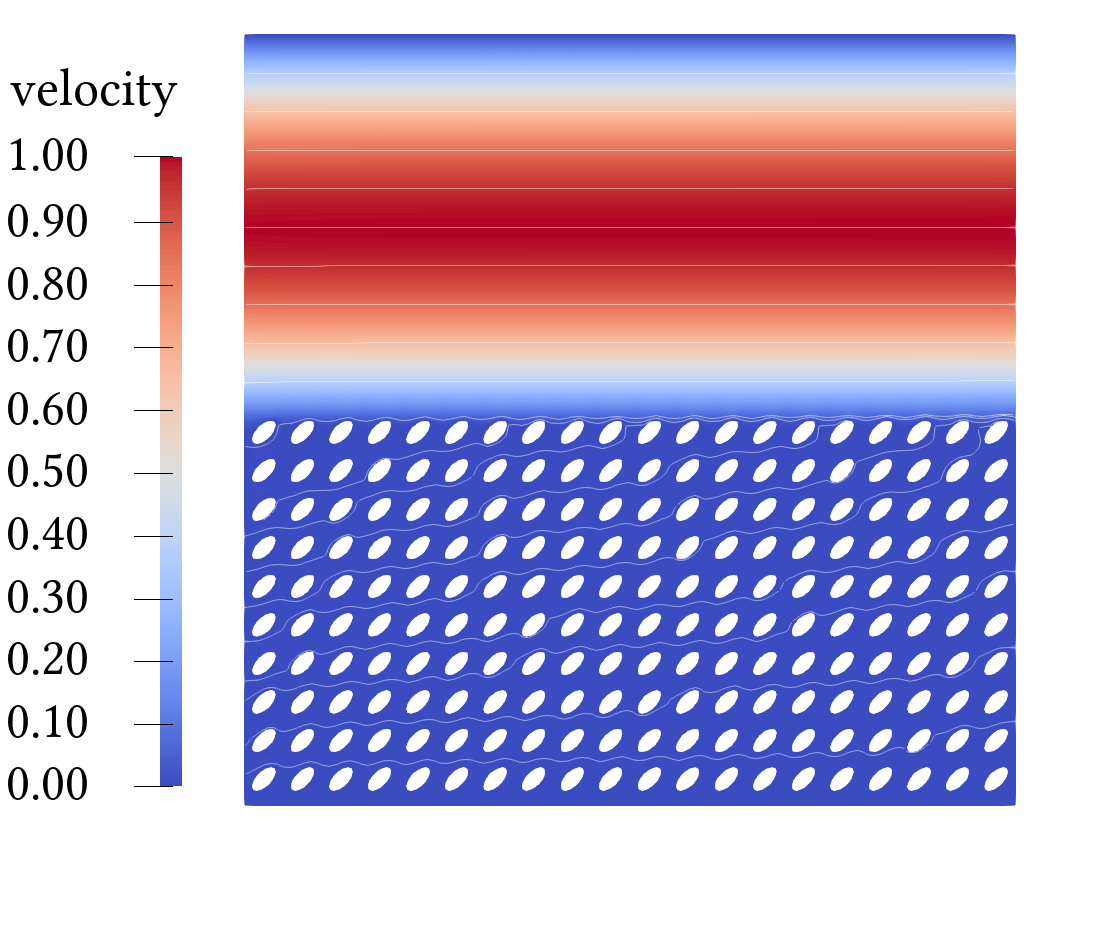} 
\includegraphics[scale=0.66]{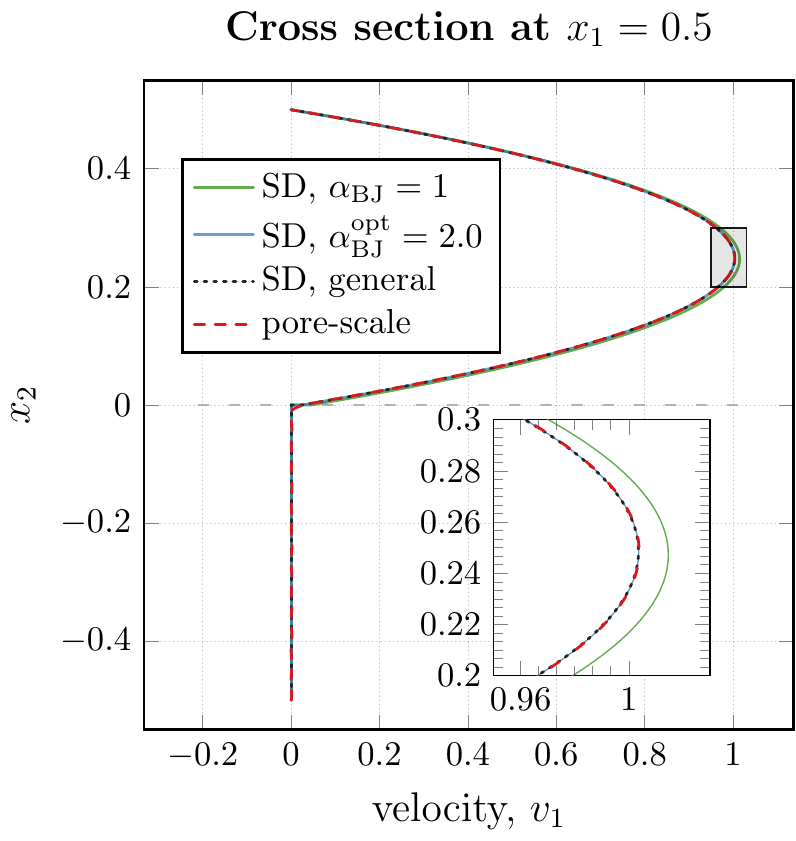}
\caption{Microscale velocity field (left) and tangential velocity profiles (right) for the pressure-driven flow and geometry $G_5$}  
\label{fig:pd-G5}
\end{figure}

We obtain the same optimal value of the Beavers--Joseph slip coefficient $\alpha_\BJ^\text{opt}=2.0$ (Tab.~\ref{tab:alpha-values}) for both porous-medium geometrical configurations~$G_5$ and~$G_6$. 
Therefore, we only present the simulation results for geometry $G_5$ (Fig.~\ref{fig:pd-G5}).
As for geometries $G_1$ to $G_4$, we observe that the classical conditions with the optimal Beavers--Joseph parameter~$\aopt=2.0$ or the generalized interface conditions lead to a more accurate Stokes--Darcy model than the classical coupling concept with the typically used value~$\alpha_\BJ =1$.

In~Tab.~\ref{tab:alpha-values}, we summarize the optimal Beavers--Joseph parameters $\alpha_\BJ^\text{opt}$ for all considered geometrical configurations in case of pressure-driven flow. We also present the relative errors~\eqref{eq:min} for the tangential velocity $v_1$ in the middle of the domain at $x_1=0.5$ for the classical interface conditions with two Beavers--Joseph parameters $\alpha_\BJ = \aopt$ (error $\epsilon_{v_1,\, 0.5, \, \aopt}$) and $\alpha_\BJ=1$ (error $\epsilon_{v_1,\, 0.5, \, 1}$) and for the generalized conditions (error $\epsilon_{v_1,\, 0.5}$). 


\begin{table}[!h]
    \centering
    \caption{Optimal Beavers--Joseph parameters and relative errors for the pressure-driven flow and different porous-medium geometries}
    \begin{tabular}{lcccc}\hline \\[-1.75ex]
    Geometry     &
    $\alpha_\BJ^{\mathrm{opt}}$ & $\epsilon_{v_1,\, 0.5, \, \aopt}$ 
    & $\epsilon_{v_1,\, 0.5, \, 1}$ 
    & 
    $\epsilon_{v_1,\, 0.5}$ 
    \\[1.5ex] \hline 
    \\[-2ex]
    $G_1$ 
    & 2.8 & 3.401e$-$3 & 2.794e$-$2 & 3.414e$-$3 \\[0.5ex]
    $G_2$  & 0.5 & 2.838e$-$3 & 6.941e$-$3 & 2.664e$-$3 \\[0.5ex]
    $G_3$  & 7.1 & 2.632e$-$3 & 3.758e$-$2 & 2.640e$-$3 \\[0.5ex]
    $G_4$,  $\sqrt{K}=\sqrt{k_{11}}$  & 3.0 & 4.092e$-$3 & 3.089e$-$2 & 3.771e$-$3 
    \\[0.5ex]
    $G_4$,  $\sqrt{K} = \sqrt{(k_{11}+k_{22})/2}$ & 2.8 & 4.098e$-$3 & 2.844e$-$2 & 3.771e$-$3
    \\[0.5ex]
    $G_5$  & 2.0 & 2.901e$-$3 & 1.738e$-$2 & 2.900e$-$3
    \\[0.5ex]
    $G_6$  & 2.0 & 2.901e$-$3 & 1.738e$-$2 & 2.900e$-$3

    \\[0.5ex] \hline 
    \\[0.25ex]
    \end{tabular}
    \label{tab:alpha-values}
\end{table}

To summarize, the difference between the pore-scale resolved and the macroscale simulations for parallel flows to the interface can be significantly reduced (Tab.~\ref{tab:alpha-values}) by choosing the correct value of the Beavers--Joseph slip coefficient in the classical interface conditions~\eqref{eq:IC-mass}--\eqref{eq:IC-BJ}. Alternatively, one can use the generalized conditions~\eqref{eq:ER-mass},~\eqref{eq:ER-momentum} and~\eqref{eq:IC-BJ-mod}.

\subsection{General filtration problem}
\label{sec:arbitrary}

In this section, we study a general filtration problem, where the flow is arbitrary to the porous bed. We  show that the Beavers--Joseph condition~\eqref{eq:IC-BJ} is not applicable for multidimensional flows to the fluid--porous interface and, therefore, the modified form~\eqref{eq:IC-BJ-mod} with the corresponding effective coefficients should be applied.  We consider the free-flow region $\Omega_\FF=[0,1]\times [0,0.5]$, the interface $\Gamma=[0,1]\times \{0\}$ and the porous medium $\Omega_\PM=[0,1]\times [-0.5,0]$, which includes $20 \times 10 $ circular solid grains ($\ell=1/20$) with radius $r=0.25\ell$ (geometry $G_1$). In this case, the interface conditions \eqref{eq:IC-mass}--\eqref{eq:IC-momentum} and \eqref{eq:ER-mass}--\eqref{eq:ER-momentum} coincide, since $N_s^\bl=0$ (Tab.~\ref{tab:geoms}). Therefore, the classical set of coupling conditions and the generalized one differ in the condition on the tangential velocity component: the Beavers--Joseph condition \eqref{eq:IC-BJ} and its generalization \eqref{eq:ER-tangential}, which can be written in form of~\eqref{eq:IC-BJ-mod}.

\begin{figure}[t]
    \includegraphics[scale=0.16]{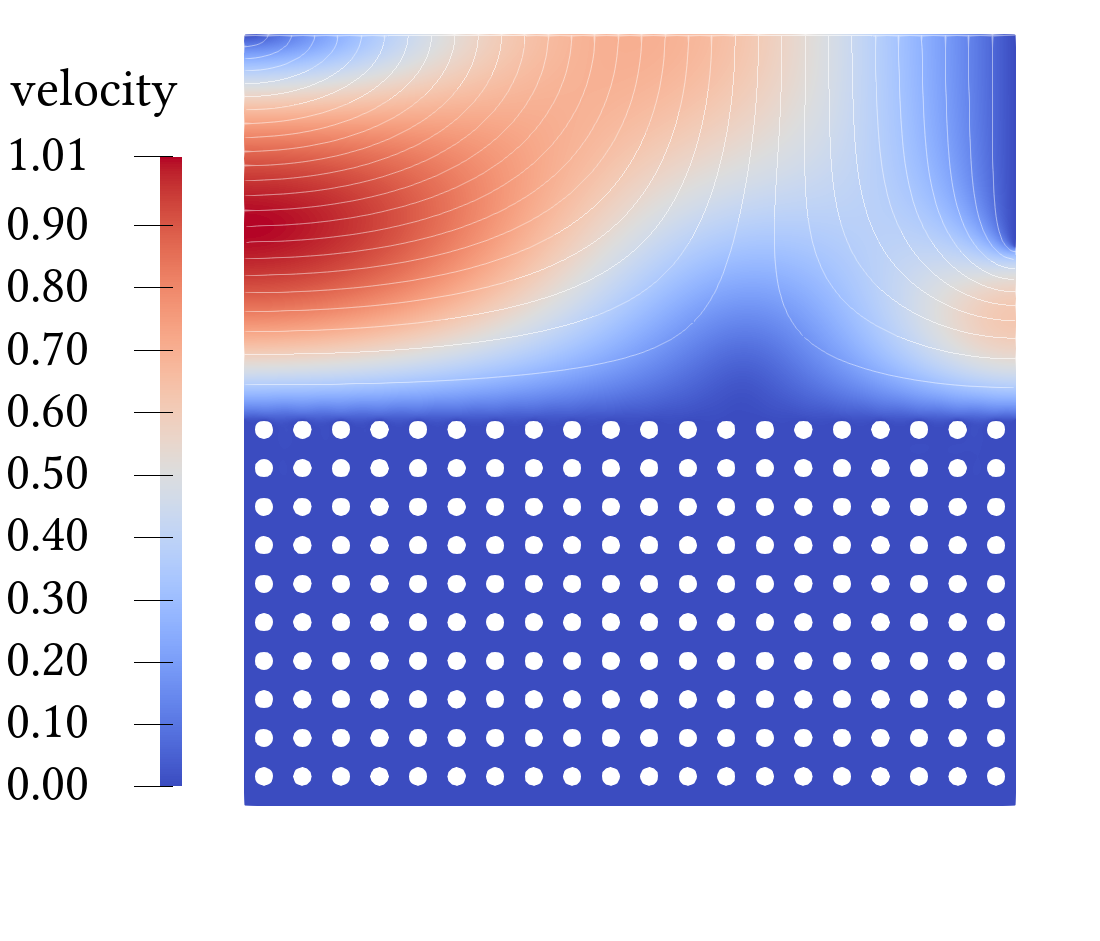} 
    \includegraphics[scale=0.66]{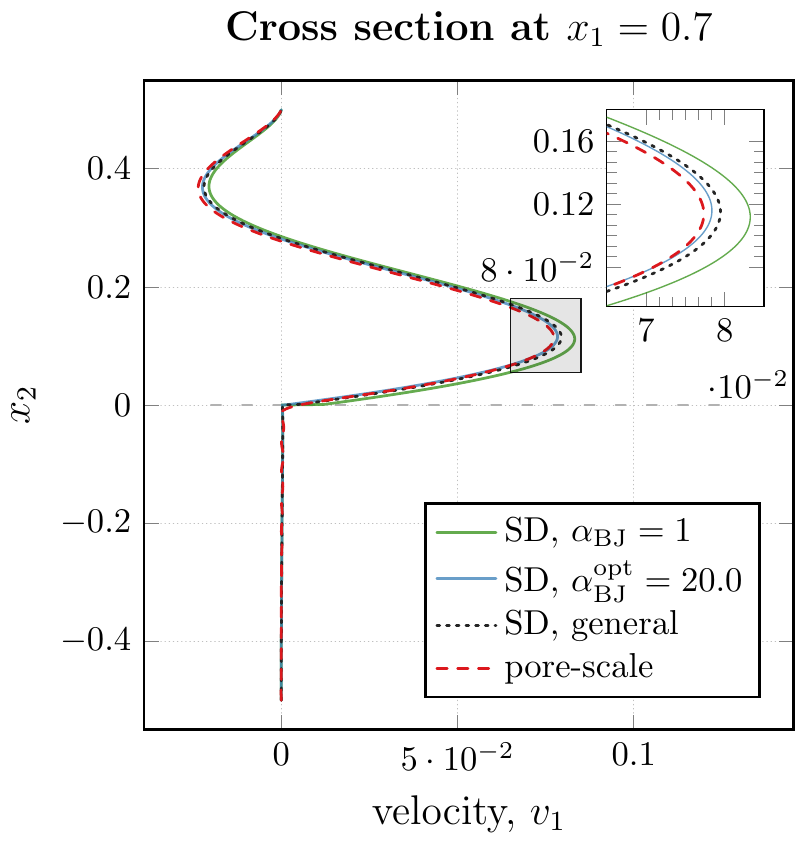}
    \caption{Microscale velocity field (left) and tangential velocity profiles (right) for the general filtration problem with geometry $G_1$ and $\ell=1/20$}
    \label{fig:general}
\end{figure}

For the pore-scale problem~\eqref{eq:micro}--\eqref{eq:micro-BC}, we impose the following boundary conditions to obtain arbitrary flow to the porous layer (Fig.~\ref{fig:general}, left):
\begin{equation}\label{eq:general-BC-micro}
\begin{split}
    \overline{\vec{v}} &= (0,-0.7\operatorname{sin}(\pi x_1)) \text{\; on \;}\Gamma_{D}^\text{in}\, , \qquad \
    \quad \overline{\vec{v}} = \vec 0 \text{\; on \;}\Gamma_{D}^\text{wall} \, ,
    \quad\
    \\
    \overline{\vec h} \vdot \vec n &=  0  \hspace{16.5ex}\text{\; on \;} \Gamma^\text{out}\, , \qquad 
    \overline{\vec{v}} \vdot \vec \tau = 0 \text{\; on \;} \Gamma^\text{out}\, ,
    \end{split}
\end{equation}
where $\Gamma_{D}^\text{in}= [0,1] \times \{0.5\}$, $\Gamma^\text{out} = \left(\{0\} \times [0,0.5] \right) \cup \left( \{1\} \times [0,0.225] \right)$, and $\Gamma_{D}^\text{wall} = \partial \Omega \setminus \left( \Gamma_{D}^\text{in} \cup \Gamma^\text{out}\right)$. 

The macroscale Stokes/Darcy model~\eqref{eq:FF}--\eqref{eq:PM-BC} with the classical interface conditions \eqref{eq:IC-mass}--\eqref{eq:IC-BJ} or the generalized conditions \eqref{eq:ER-mass}--\eqref{eq:ER-tangential} is complemented by the following boundary conditions
\begin{equation}\label{eq:general-BC-macro}
\begin{split}
    \overline{\vec{v}} &=  (0,-0.7\operatorname{sin}(\pi x_1)) \text{\; on \; }\Gamma_{D}^\text{in} \, ,
    \qquad \
    \overline{\vec{v}} = \vec 0 \text{\; on \; }\Gamma_{D,\FF}^\text{wall} \, ,
    \\
    \overline{\vec h} \vdot \vec n &= 0, \quad 
    \overline{\vec{v}} \vdot \vec \tau = 0 \qquad \, \text{\; on \;} \Gamma^\text{out}\, ,
    \qquad \
    \overline v = 0 \text{\; on \; }\Gamma_{N,\PM}^\text{wall} \, ,
    \end{split}
\end{equation}
where $\Gamma_{D,\FF}^\text{wall} = \Gamma_{D}^\text{wall} \cap \partial \Omega_\FF$ and $\Gamma_{N,\PM}^\text{wall} = \Gamma_{D}^\text{wall} \cap \partial \Omega_\PM$. 

In~Fig.~\ref{fig:general} (left), we present the pore-scale velocity field for the general filtration problem with arbitrary flow direction to the porous layer. First, we try to determine an optimal value of the Beavers--Joseph parameter in condition~\eqref{eq:IC-BJ} in the same way as in~Section~\ref{sec:pressure-driven}. Since the flow is non-parallel to the fluid--porous interface in this case, we consider a broader range for the Beavers--Joseph parameter setting the upper boundary $M=100$ in \algo~\ref{Algorithm2}. 

In~Fig.~\ref{fig:general} (right), we provide the pore-scale resolved and macroscale velocity profiles for the tangential component  at $x_1=0.7$, where the flow direction is arbitrary to the interface. The macroscale tangential velocity profile computed using the typical value $\alpha_\BJ =1$ does not fit to the pore-scale resolved velocity. We noticed that the difference between the macroscale and the pore-scale simulation results for the tangential velocity~$\epsilon_{v_1, \, 0.7, \, \alpha_\BJ}$ given by~\eqref{eq:min} is getting smaller for bigger values of the Beavers--Joseph slip coefficient $\alpha_\BJ$, however, these improvements are minor for $\alpha_\BJ \in [20,100]$. Therefore, we take $\alpha_\BJ^\text{opt}=20$ at $x_1 = 0.7$ for the tangential velocity. As can be seen from~Fig.~\ref{fig:general} (right), the macroscale profile for the Beavers--Joseph slip coefficient $\alpha_\BJ=20$ fits much better to the pore-scale one than for the traditionally used value $\alpha_\BJ=1$.

Since the flow is arbitrary to the interface, we consider four different cross sections and provide the optimal values~$\alpha_\BJ^\text{opt}$ and the errors~\eqref{eq:min} for both velocity components~$v_1$ and~$v_2$ in~Tab.~\ref{tab:general}. 
The microscale surface roughness, porosity and permeability reflected in the Beavers--Joseph parameter $\alpha_\BJ$ do not change along the fluid--porous interface for the considered porous media, therefore, the slip coefficient~$\alpha_\BJ$ must be constant along the interface.
However, we observe that the optimal value of the Beavers--Joseph parameter is different for every cross section and each velocity component.  
This fact indicates the unsuitability of the Beavers--Joseph condition~\eqref{eq:IC-BJ} for arbitrary flows. 

\begin{table}[!ht]
    \centering
        \caption{Optimal Beavers--Joseph parameters and relative errors for the general filtration problem and geometry~$G_1$}
    \begin{tabular}{l c c c c c}
    \hline \\[-1.5ex]
    Cross section & Velocity & $\aopt$ & $\epsilon_{v_i,\, x_1, \, \aopt}$ &
    $\epsilon_{v_i,\, x_1, \, 1}$ & $\epsilon_{v_i,\, x_1}$ \\[1.5ex]
    \hline 
    \\[-2.5ex]
    $x_1 = 0.5$ & $v_1$ & 5.0 & 6.620e$-$3 & 2.484e$-$2 & 7.917e$-$3
    \\[0.5ex] 
    $x_1 = 0.5$ & $v_2$ & 3.9 & 1.283e$-$3  & 8.858e$-$3  & 1.248e$-$3
    \\[0.5ex] \hline 
    \\[-2.5ex]
    $x_1 = 0.7$ & $v_1$ & 20.0 & 3.468e$-$2 & 9.901e$-$2 &  4.676e$-$2
    \\[0.5ex] 
    $x_1 = 0.7$ & $v_2$ & 3.3 & 1.638e$-$3  & 1.087e$-$2 &  1.764e$-$3
    \\[0.5ex]     \hline 
    \\[-2.5ex]
    $x_1 = 0.8$ & $v_1$ & 3.7 & 1.510e$-$2 & 4.039e$-$2 &  1.502e$-$2
    \\[0.5ex] 
    $x_1 = 0.8$ & $v_2$ & 4.1 & 1.691e$-$3  & 1.108e$-$2 & 2.239e$-$3
    \\[0.5ex]  \hline 
    \\[-2.5ex]
    $x_1 = 0.9$ & $v_1$ & 3.2 & 1.215e$-$2 & 3.529e$-$2 & 1.179e$-$2
    \\[0.5ex] 
    $x_1 = 0.9$ & $v_2$ & 3.6 & 4.638e$-$3  & 9.416e$-$3  &  4.910e$-$3
    \\[0.5ex]
    \hline
    \\[0.25ex]
    \end{tabular}
    \label{tab:general}
\end{table}

In contrast, when the generalized condition~\eqref{eq:ER-tangential} or, equivalently,~\eqref{eq:IC-BJ-mod} is used instead of condition~\eqref{eq:IC-BJ}, the pore-scale and the macroscale simulations fit very well (Fig.~\ref{fig:general} and Tab.~\ref{tab:general}). 
Notice that the errors for the generalized coupling conditions are of the same order as for the classical conditions with the optimal value of the Beavers--Joseph parameter determined at every cross-section (Tab.~\ref{tab:general}). In case of the generalized conditions all effective coefficients are computed only once. This is a huge advantage in terms of the computational effort.   
Therefore, the generalized coupling condition on the tangential velocity should be used. We advice the researchers having their codes with the Beavers--Joseph condition to update only two parameters appearing in the generalized condition in form of \eqref{eq:IC-BJ-mod}.


\section{Conclusions}\label{sec:conclusion}\label{sec:conclusions}

In this paper, we analyzed the applicability of the Beavers--Joseph interface condition~\eqref{eq:IC-BJ} to different flow problems, developed an efficient two-level algorithm to determine the slip coefficient $\alpha_\BJ$ and proposed a  modification of the Beavers--Joseph interface condition, which is suitable for arbitrary flow directions. 
We found out that (i)~the Beavers--Joseph parameter is in general not constant along the fluid--porous interface and thus the Beavers--Joseph condition is unsuitable for arbitrary flow directions; (ii)~the typically used value $\alpha_\BJ=1$ is not correct for many flow problems; (iii)~the optimal value of the Beavers--Joseph slip coefficient can be found only for unidirectional flows;
(iv) the slip coefficient~$\alpha_\BJ$ depends on the pore-scale geometrical information including the microscale surface roughness; (v)~the generalized interface condition~\eqref{eq:IC-BJ-mod} should be used instead of the Beavers--Joseph condition~\eqref{eq:IC-BJ} for arbitrary flows. These conclusions are made based on the detailed study of two flow problems: the pressure-driven flow parallel to the fluid--porous interface and the general filtration problem, where the fluid flow is arbitrary to the porous bed.

We analyzed different types of porous media: isotropic, orthotropic, and anisotropic. To find the optimal value of the Beavers--Joseph parameter $\alpha_\BJ$ we propose an efficient two-level numerical algorithm based on minimizing the difference between the macroscale solution of the Stokes/Darcy problem and the pore-scale resolved solution. In order to study the dependency of the optimal Beavers--Joseph parameter on the pore geometry, we constructed porous media having the same effective properties (porosity, permeability) but different pore morphologies (circular and squared solid grains) and considered different configurations of the porous bed having the same microscale surface roughness.

For parallel flows to the porous medium, the Beavers--Joseph slip coefficient can be determined using pore-scale information that leads to accurate simulation results for the corresponding coupled Stokes/Darcy problem. However, it is not possible to find an optimal value $\alpha_\BJ$ for general filtration  problems with arbitrary flow directions to the fluid--porous interface. Therefore, the Beavers--Joseph condition is not applicable for such flow problems. In the case of arbitrary flow directions, alternative sets of interface conditions should be used,  e.g. those proposed in \citep{Angot_etal_17, Eggenweiler_Rybak_MMS20,Sudhakar_21}. We reformulated the generalized interface condition on the tangential velocity proposed in \citep{Eggenweiler_Rybak_MMS20} in the form~\eqref{eq:IC-BJ-mod} of the Beavers--Joseph condition~\eqref{eq:IC-BJ}. This modification does not contain any unknown parameters and it is suitable for arbitrary flows in the Stokes/Darcy systems.


\bmhead{Author contributions} All authors contributed to the study conception and design. The two-level algorithm was proposed and implemented by Paula Strohbeck. The code for macroscale problems was developed by Elissa Eggenweiler and Iryna Rybak. Effective model parameters and microscale solutions were computed by Elissa Eggenweiler. The first draft of the manuscript was written by Iryna Rybak and Elissa Eggenweiler, and all authors commented on previous versions of the manuscript. All authors read and approved the final manuscript.

\bmhead{Funding}
The work is funded by the Deutsche Forschungsgemeinschaft (DFG, German Research Foundation) -- Project Number 327154368 -- SFB 1313.

\bmhead{Data availability} The datasets generated and analyzed during the current study are available from
the corresponding author on reasonable request.

\subsection*{Declarations}

\textbf{Conflict of interest.}
The authors have no relevant financial or non-financial interests to disclose.

\bibliography{CaF}

\end{document}